\documentclass{iopart}

\usepackage[comma,numbers,sort&compress]{natbib}
\usepackage{iopams}
\usepackage{setstack}
\usepackage{amsfonts}
\usepackage{amssymb}

\usepackage{multirow}
\usepackage{color}
\usepackage{colortbl}
\usepackage{setspace}

\usepackage{epsfig}

\def\newblock{\hskip .11em plus .33em minus .07em}

\newcommand{\code}[1]{\texttt{#1}}

\def\mev{\, \mathrm{MeV}}

\def\apj{Astrophys. J.}
\def\apjl{Astrophys. J. Lett.}
\def\apjs{Astrophys. J. Supp. Ser. }

\def\aap{Astron. Astrophys. }
\def\aaps{Astron. Astrophys. Suppl.}

\def\physrep{Phys. Rep. }
\def\mnras{Mon. Not. Roy. Astron. Soc. }

\def\prl{Phys. Rev. Lett.}
\def\prd{Phys. Rev. D.}

\begin{document}

\title[A New Open-Source Code for 1D Stellar Collapse]
{A New Open-Source Code for Spherically-Symmetric Stellar Collapse to\\
Neutron Stars and Black Holes}

\author{Evan O'Connor}
\ead{evanoc@tapir.caltech.edu}

\address{\scriptsize 
  TAPIR, Mail Code 350-17,\\
  California Institute of Technology,
  Pasadena, California 91125, USA}

\author{Christian D Ott}
\ead{cott@tapir.caltech.edu}

\address{\scriptsize 
  TAPIR, Mail Code 350-17,\\
  California Institute of Technology,
  Pasadena, California 91125, USA \\
  and \\ 
  Niels Bohr International Academy, The Niels Bohr Institute,\\
  Copenhagen, Denmark 
  \\
  and \\
  Center for Computation and Technology,
  Louisiana State University,\\ Baton Rouge, LA, USA}

%%%%%%%%%%%%%%%%%%%%%%%%%%%%%%%%%%%%%%%%%%%%%%%%%%%%%%%%%%%%%%%%%%%%%%%%
%%%%%%%%%%%%%%%%%%%%%%%%%%%%%%%%%%%%%%%%%%%%%%%%%%%%%%%%%%%%%%%%%%%%%%%%
%%%%%%%%%%%%%%%%%%%%%%%%%%%%%%%%%%%%%%%%%%%%%%%%%%%%%%%%%%%%%%%%%%%%%%%%
\begin{abstract}
  We present the new open-source spherically-symmetric
  general-relativistic (GR) hydrodynamics code \code{GR1D}. It is
  based on the Eulerian formulation of GR hydrodynamics (GRHD) put
  forth by Romero-Ib\'a\~ nez-Gourgoulhon and employs radial-gauge,
  polar-slicing coordinates in which the 3+1 equations simplify
  substantially. We discretize the GRHD equations with a finite-volume
  scheme, employing piecewise-parabolic reconstruction and an
  approximate Riemann solver. \code{GR1D} is intended for the
  simulation of stellar collapse to neutron stars and black holes and
  will also serve as a testbed for modeling technology to be
  incorporated in multi-D GR codes. Its GRHD part is coupled to
  various finite-temperature microphysical equations of state in
  tabulated form that we make available with \code{GR1D}.  An
  approximate deleptonization scheme for the collapse phase and a
  neutrino-leakage/heating scheme for the postbounce epoch are
  included and described. We also derive the equations for effective
  rotation in 1D and implement them in \code{GR1D}. We present an
  array of standard test calculations and also show how simple
  analytic equations of state in combination with presupernova models
  from stellar evolutionary calculations can be used to study 
  qualitative aspects of black hole formation in failing
  \emph{rotating} core-collapse supernovae. In addition, we present a
  simulation with microphysical EOS and neutrino leakage/heating of a
  failing core-collapse supernova and black hole formation in a
  presupernova model of a $40$-$M_\odot$ zero-age main-sequence
  star. We find good agreement on the time of black hole formation
  (within 20\%) and last stable protoneutron star mass (within 10\%) with
  predictions from simulations with full Boltzmann neutrino radiation
  hydrodynamics.
\end{abstract}


%%%%%%%%%%%%%%%%%%%%%%%%%%%%%%%%%%%%%%%%%%%%%%%%%%%%%%%%%%%%%%%%%%%%%%%%
%%%%%%%%%%%%%%%%%%%%%%%%%%%%%%%%%%%%%%%%%%%%%%%%%%%%%%%%%%%%%%%%%%%%%%%%
%%%%%%%%%%%%%%%%%%%%%%%%%%%%%%%%%%%%%%%%%%%%%%%%%%%%%%%%%%%%%%%%%%%%%%%%

\section{Introduction}
\label{section:intro}

Stellar core collapse is among the most energetic phenomena in the
modern universe and liberates of the order of a few hundred [B]ethe
($1\, \mathrm{B} = 10^{51}\,\mathrm{erg}$) of gravitational energy as
the core of a massive star (zero-age main-sequence [ZAMS] mass $
8-10\,M_\odot \lesssim M \lesssim 100\,M_\odot$) is compressed from a
radius of $\sim 1500\,\mathrm{km}$ and central density $\rho_c \sim
10^{10}\,\mathrm{g\,cm}^{-3}$ to $\sim 15\,\mathrm{km}$ and $\rho_c$
in excess of nuclear density. Most ($\sim 99\%$) of this energy is
ultimately radiated in neutrinos, but a small fraction ($\sim 1
\,\mathrm{B}$) may be converted into kinetic and internal energy of an
outgoing shock wave and may result in a core-collapse supernova
explosion within the first seconds after collapse. The precise mode of
conversion, the \emph{core-collapse supernova mechanism}, is uncertain
and has been the enigma of supernova theory for the past five decades
(e.g.,
\cite{bethe:90,liebendoerfer:05,buras:06b,marek:09,janka:07,burrows:07bethe,bruenn:09,ott:09b}). At
the densities and velocities encountered in stellar collapse, the
inclusion of general relativistic effects is not an optional model
sophistication, but a necessity for quantitatively and qualitatively
reliable results.  Importantly, general relativity (GR) predicts that
the protoneutron star (PNS) formed in the initial collapse will
undergo a second gravitational instability and collapse to a black
hole (BH), if continued accretion pushes it over the maximum mass
supported by the strong force and nucleon degeneracy.  This may happen
if the supernova mechanism fails and no explosion is launched or due
to fallback accretion if an explosion occurs, but is too weak to
unbind the entire stellar envelope \cite{zhang:08}.  In both cases,
and provided sufficient angular momentum and its appropriate
distribution in the progenitor star, the newly formed \emph{collapsar}
may become the central engine for a long-soft gamma-ray burst (GRB)
\cite{woosley:93,wb:06}.

General relativistic computational models of stellar collapse have a
long pedigree, starting with the spherically-symmetric (1D) Lagrangian
work of May \& White in the mid-1960s~\cite{may:66}, based on the
comoving GR hydrodynamics formulation in orthogonal coordinates by
Misner \& Sharp \cite{misner:64} and using a finite-difference scheme
with an artificial viscosity \cite{vonneumann:50} approach to handle
shocks.  Much subsequent 1D GR work \cite{vanriper:79,baumgarte:95,
  swesty:95,liebendoerfer:02,miralles:91,schinder:88} was based on
this or similar approaches, including full radiation-hydrodynamics
stellar collapse and core-collapse supernova simulations with
finite-temperature microphysical equations of state (EOS)
\cite{wilson:71,bruenn:85,baron:85,baron:89,liebendoerfer:04}.
Eulerian formulations, more suited for extension to multi-D simulations,
were introduced later and used maximal slicing
\cite{wilson:79,shapiro:79,shapiro:80,mezzacappa:89}, or radial-gauge,
polar-slicing (RGPS) \cite{gourgoulhon:91}.  These schemes, with the
exception of \cite{gourgoulhon:91}, who employed pseudospectral
methods, still used artificial viscosity approaches to shock
treatment. More accurate, high-resolution shock-capturing (HRSC)
approaches to GR stellar collapse based on higher-order Gudonov
schemes and Riemann solvers were introduced by
Marti~et~al.~\cite{marti:90} and Yamada \cite{yamada:97} in the
Lagrangian context, by Marti~et~al.~\cite{marti:91} in the
fixed-background Eulerian case, and by Romero~et~al.~\cite{romero:96}
and Noble \cite{noble:03phd} in the RGPS Eulerian frame.  Yamada's
approach was later extended to include microphysical EOS and radiation
transport \cite{yamada:99,sumi:05}. Gourgoulhon \& Haensel
\cite{gourgoulhon:93} included an approximate neutrino transport
treatment in their code.  Preliminary results of Romero's code with 
a microphysical EOS and a neutrino leakage scheme were published in
\cite{romero:97,pons:97}.

State-of-the-art simulations of stellar collapse and of the postbounce
supernova evolution strongly suggest that multi-D dynamics is crucial
for the core-collapse supernova mechanism to succeed in massive stars
(e.g., \cite{liebendoerfer:05,thompson:03,marek:09,
  burrows:06,burrows:07b,ott:08}). Present {\mbox{multi-D}}
core-collapse supernova codes are either Newtonian
\cite{swesty:09,burrows:07a, ott:08} or employ Newtonian dynamics with
relativistic corrections to the gravitational potential
\cite{buras:06b,marek:09, bruenn:09}. Multi-D simulations in
conformally-flat \cite{isenberg:08} or full GR traditionally relied on
simple analytic EOS and polytropic initial models and neglected
crucial neutrino effects (see, e.g.,
\cite{dimmelmeier:02,dimmelmeier:05,shibata:04,shibata:05}).  Only
recently have the first axisymmetric (2D)
\cite{dimmelmeier:07,dimmelmeier:08} and 3D \cite{ott:07prl,ott:07cqg}
GR core collapse simulations become available that employ
microphysical EOS and an approximate treatment of deleptonization in
the collapse phase, but postbounce neutrino transport, cooling, and
heating are still not taken into account in these models. However,
very recently, M\"uller \cite{mueller:09phd} has succeeded in
implementing the complex and computationally-intensive
radiation-transport scheme of \cite{buras:06a} in the 2D
conformally-flat GR framework of \cite{dimmelmeier:02,dimmelmeier:05}
and first results are forthcoming \cite{mueller:10}.

In this article, we lay the foundations for a new and open approach to
the stellar collapse and core-collapse supernova problem in GR.  We
discuss the formulation and implementation of the code \code{GR1D}, a
new, spherically-symmetric Eulerian GR code for stellar collapse to
neutron stars and black holes with approximate pre- and postbounce
neutrino treatment. We release \code{GR1D} and all its microphysics
and input physics as open source to be downloaded from {\tt
  http://www.stellarcollapse.org}. It is meant to complement
open-source 3D GR codes such as \code{Whisky}~\cite{whiskyweb} that do
not come with microphysics and neutrino approximations. At the same
time, we intend \code{GR1D} to serve as an efficient 1D GR testbed for
new modeling technology to be eventually incorporated in multi-D
codes. In addition, \code{GR1D} and its microphysics components can
readily be adapted for use in the computational modeling of problems
involving some or much of the same physics as in the stellar collapse
problem, e.g., the postmerger phase of double neutron-star or
black-hole -- neutron-star coalescence.

We base \code{GR1D} on the conceptually simple and computationally
efficient RGPS formalism of \cite{gourgoulhon:91}. \code{GR1D}, like
the code of \cite{romero:96}, employs a Eulerian formulation of GR
hydrodynamics with HRSC and works on non-equidistant grids. For the
first time in the 1D GR context, we derive and implement in
\code{GR1D} an extension of the 1D GR hydrodynamics equations to
include rotation in an effective fashion.  For completeness and
comparison of Newtonian and GR dynamics, \code{GR1D} also implements
1D Newtonian hydrodynamics.  \code{GR1D} operates with analytic EOS as
well as with tabulated microphysical EOS through a general EOS
interface. We discuss and provide EOS tables for the EOS of
Lattimer-Swesty \cite{lseos:91} and the one of
H.~Shen~et~al.~\cite{shen:98a,shen:98b}. Furthermore, we discuss and
include in \code{GR1D} the deleptonization treatment of
\cite{liebendoerfer:05fakenu} for the collapse phase and a postbounce
3-flavor neutrino treatment based on the leakage schemes of
\cite{ruffert:96,rosswog:03b} as well as an approximate way of
including neutrino heating. 

Due to these approximations in the neutrino treatment, \code{GR1D} in
its present form cannot be used for accurate simulations addressing
the core-collapse supernova mechanism or neutrino-induced
nucleosynthesis. However, we find that with the present treatment,
\code{GR1D} reproduces very well qualitatively the salient features of
the postbounce evolution of core-collapse supernovae as predicted by
full 1D radiation-hydrodynamics simulations. Moreover, we find that
\code{GR1D} may be used to make quantitatively reliable predictions on
the time of black hole formation in failing core-collapse supernovae
and on the maximum mass of the PNS.

This article is structured as follows. In \sref{sec:curvehydro}, we
discuss our 1D GR hydrodynamics and curvature equations and their
implementation in \code{GR1D}.  \Sref{sec:EOS} introduces the EOS
provided with \code{GR1D} and in \sref{sec:leakage} we detail our
pre-bounce deleptonization and postbounce leakage and neutrino heating
schemes. A number of code tests and example simulations are presented
in \sref{sec:tests} and \ref{sec:sampleresults}.  We wrap up and conclude
in \sref{sec:summary}. 

We assume spacelike signature $(-,+,+,+)$ and, unless mentioned
otherwise, use units of $G = c = M_\odot = 1$, but use cgs units
for the microphysics and neutrino leakage/heating quantities.

\section{1D GR Hydrodynamics and Curvature Equations}
\label{sec:curvehydro}

\subsection{Curvature Equations in 1D RGPS}
\label{sec:curve}

We follow \cite{romero:96,gourgoulhon:91} who formulate the $3+1$ GR
curvature and hydrodynamics equations in RGPS coordinates. In these
coordinates and in spherical symmetry, the shift vector vanishes and
the metric is diagonal and closely resembles the Schwarzschild
metric. The invariant line element is
\begin{eqnarray}
\nonumber
ds^2 & = & g_{\mu \nu} dx^\mu dx^\nu\,\,, \\
& = & -\alpha(r,t)^2 dt^2 + X(r,t)^2 dr^2 + r^2 d\Omega^2 \,\,,
\end{eqnarray}
\noindent 
where $\alpha$ and $X$ can be written more conveniently as functions of
a metric potential, $\Phi(r,t)$, and the enclosed gravitational
mass $M_\mathrm{grav}(r,t) = m(r,t)$,
\begin{equation}
\alpha(r,t) = \exp\left[\Phi(r,t)\right],\hspace*{1cm}
X(r,t) = \left( 1 - {2 m(r,t) \over
    r}\right)^{-1/2}\,\,.
\label{eq:metriccoefficients}
\end{equation}
We assume ideal hydrodynamics for which the fluid stress-energy tensor
 and the matter current density are-
\begin{equation}
T^{\mu \nu} = \rho h u^\mu u^\nu + Pg^{\mu \nu}\, \,\,\,
\mathrm{and}\, \, \, \, 
J^\mu = \rho u^\mu \,,
\end{equation}
\noindent
where $\rho$ is the baryonic density, $P$ is the fluid pressure, $h$
is the specific enthalpy equal to $1 + \epsilon + P/\rho$ with
$\epsilon$ being the specific internal energy.  $u^\mu$ is the four-velocity
and, in 1D without rotation, is equal to $[W/\alpha,Wv^r,0,0]$. 
$W=\left(1-v^2\right)^{-1/2}$ is the Lorentz factor and $v = Xv^r$.
The equation for the gravitational mass needed for determining the
metric coefficient $X(r,t)$ of \eref{eq:metriccoefficients} is
derived from the Hamiltonian constraint equation and reads
\begin{equation}
m(r,t) = 4\pi \int_0^r (\rho h W^2 - P + \tau^\nu_m) {r^\prime}^2 dr^\prime\,\,.
\label{eq:mgrav}
\end{equation}
Here, $\tau^\nu_m$ is the contribution to the gravitational mass from
the energy and pressure of trapped neutrinos (see \sref{sec:neutrinopressure}).
The expression for the metric potential $\Phi(r,t)$ is determined
via the momentum constraints, taking into account the polar slicing condition
that imposes $\tr K = K_r^{\,\,r}$, where $K_{ij}$ is the extrinsic curvature
tensor (see \cite{gourgoulhon:91,noble:03phd} for details).  It reads,
\begin{equation}
 \Phi(r,t) = \int_0^rX^2\left[{m(r^\prime,t) \over {r^\prime}^2} +
   4\pi r^\prime(\rho h W^2
   v^2+ P + \tau^\nu_\Phi)\right]dr^\prime + \Phi_0\,\,,
\label{eq:phi}
\end{equation}
\noindent
where analogous to \eref{eq:mgrav}, $\tau^\nu_\Phi$ accounts for the effect of
trapped neutrinos. $\Phi_0$ is determined by matching the solution
at the star's surface ($r=R_\star$) to the Schwarzschild metric,
\begin{equation}
\Phi(R_\star,t) =\ln\left[\alpha(R_\star,t)\right] =
{1 \over 2}\ln\left[1-{2m(R_\star,t) \over R_\star}\right].
\end{equation}

\noindent We use standard $2^{\mathrm{nd}}$~order methods to perform the integrals
in \eref{eq:mgrav} and \eref{eq:phi} and obtain values at cell centers
as well as at cell interfaces.  

\subsection{GR Hydrodynamics in 1D RGPS}
\label{sec:grhydro}

\noindent The evolution equations for the matter fields are derived
from the local conservation laws for the stress-energy tensor,
$\nabla_\mu T^{\mu\nu} = 0$, and for the matter current density
$\nabla_\mu J^\mu = 0$. We write the GR hydrodynamics equations along
the lines of the flux-conservative Valencia formulation (e.g.,
\cite{banyuls:97,font:00,font:08}) with modifications for
spherically-symmetric flows proposed by \cite{romero:96} and neutrino sources.  Derivation
details are presented in \ref{sec:appendix}.

We write the set of evolution equations as,
\begin{equation}
\partial_t \vec{U} + {1 \over r^2} \partial_r\left[{\alpha r^2  \over X}\vec{F}\right] = 
\vec{\mathcal{S}}\,,\label{eq:evolution}
\end{equation}
\noindent
where $\vec{U}$ is the set of conserved variables, $\vec{F}$ is
their flux vector, and $\vec{\mathcal{S}}$ is the vector containing
gravitational, geometric, and neutrino-matter interaction sources
and sinks. In 1D and without rotation, $\vec{U} = [D,DY_e, S^r,\tau]$.
The conserved variables are functions of the primitive variables
$\rho, Y_e, \epsilon, v,$ and $P$ and are given by
\begin{eqnarray}
\nonumber
 D & = & \alpha X J^t =  X \rho W\,\,, \\
 \nonumber
 DY_e &=& \alpha X Y_e J^t = X \rho W Y_e\,\,,\\
\nonumber
S^r & = &\alpha X T^{tr} =   \rho h W^2 v\,\,, \\
\tau & = &\alpha^2 T^{tt} - D  = \rho h W^2  - P - D\,\,,
\end{eqnarray}
where $Y_e$ is the electron fraction, the number of electrons
per baryon, and the only compositional variable needed to
describe matter in nuclear statistical equilibrium (NSE). Note that
there is a misprint in the central part of Eq.~9 of \cite{romero:96}
which is missing a factor of $X$ which we have corrected here.
The flux $\vec{F}$ is given by
$\vec{F} = [Dv,DY_ev,S^rv+P,S^r-Dv]$ and the sources and sinks  
are given by
\begin{eqnarray}
\vec{\mathcal{S}} =& \bigg[ 0 , R^\nu_{Y_e} ,(S^rv -
 \tau - D)\alpha X\left(8 \pi r P + {m \over r^2}\right) + \alpha P X
       {m \over r^2}  \nonumber \\
&+ {2 \alpha P \over X r} + Q_{S^r}^{\nu,\mathrm{E}} +
       Q_{S^r}^{\nu,\mathrm{M}}, Q_\tau^{\nu,\mathrm{E}} + Q_\tau^{\nu,\mathrm{M}}
         \bigg]\,\,.
\label{eq:sources}
\end{eqnarray}
The source and sink terms $R^\nu_{Y_e}, Q_{S^r}^{\nu,\mathrm{E}},
Q_{S^r}^{\nu,\mathrm{M}}, Q_\tau^{\nu,\mathrm{E}},$ and $Q_\tau^{\nu,\mathrm{M}}$ 
are associated with neutrinos and are discussed in \sref{sec:leakage} and
derived in \ref{sec:appendix}.

We use a semi-discrete approach and first
discretize~\eref{eq:evolution} in space, then apply the method of
lines (MoL, \cite{Hyman-1976-Courant-MOL-report}) and perform the time
integration of the conserved variables via standard $2^\mathrm{nd}$ or
$3^\mathrm{rd}$~order Runge-Kutta integrators with a Courant factor of
$0.5$.

The spatial discretization follows the finite-volume approach (e.g.,
\cite{romero:96,font:08}) and all variables are defined at cell
centers $i$ and must be reconstructed (i.e., interpolated) at cell
interfaces, where inter-cell fluxes are computed. This interpolation
must be monotonic to ensure stability. We use the nominally
$3^\mathrm{rd}$~order (in smooth parts of the flow)
piecewise-parabolic method (PPM, \cite{colella:84}) to interpolate the
primitive variables and then set up the conserved variables at the
cell interfaces. We also implement piecewise-constant reconstruction
as well as piecewise-linear (total-variation-diminishing [TVD])
reconstruction with Van Leer's limiter \cite{vanleer:77}. The latter
we use exclusively in the innermost 3 to 5 zones to avoid oscillations
near the origin.

Once the variables have been reconstructed at the cell interfaces, we
evaluate the physical interface fluxes $\vec{F}_{i+1/2}$ with the HLLE
Riemann solver \cite{HLLE:88}. The right-hand-side (RHS) flux update term
for $\vec{U}_i$ then reads,  
\begin{equation}
\hspace*{-2cm}\mathrm{RHS}_i = - \frac{1}{r_i^2 \Delta r_i} 
\left[ \frac{\alpha_{i+1/2} r_{i+1/2}^2}{X_{i+1/2}} \vec{F}_{i+1/2} - 
\frac{\alpha_{i-1/2} r_{i-1/2}^2}{X_{i-1/2}} \vec{F}_{i-1/2} 
\right]\,\,.
\end{equation}

\noindent Gravitational, geometrical, and neutrino-matter interaction
sources/sinks are not taken into account in the flux computation
and are coupled into the MoL integration.

After the update of the conserved variables $D$, $DY_e$, $S^r$ and
$\tau$, primitive variables $\rho$, $Y_e$, $v$, $\epsilon$, and
$P(\rho,\epsilon,Y_e)$ must be extracted since they are needed for the
next timestep.  In the general case, the primitive variables (with the
exception of $Y_e$) cannot be expressed algebraically in terms of the
conserved variables (see, e.g., \cite{font:00}). Hence, we employ an
iterative approach and make an initial guess using $P_\mathrm{old}$
from the previous timestep, 
\begin{equation}
\hspace*{-2cm}v\    =\    {S^r \over \tau + D + P_{old}}\,\,, \hspace*{0.3cm}
\rho\    =\    {D \over X W} \,\,, \hspace*{0.3cm} 
\epsilon\   =\  { \tau + D + P_{old}(1 - W^2) \over \rho W^2} - 1,
\end{equation}
where we note that $X$ can be calculated from the conserved
variables as $\rho h W^2 -P = \tau + D$. $W$ is calculated from the
estimate of $v$.  We then call the EOS to obtain a new
pressure and iterate this process using a Newton-Raphson method until
convergence (we typically stop the iteration at a fractional pressure
difference of $10^{-10}$ between iteration steps).

\subsection{Extension to 1.5D: Including Rotation}
\label{sec:rotation}

Lagrangian spherically-symmetric stellar evolution codes have long
included rotation and rotational effects in an approximate
fashion~(e.g., \cite{endal:78,heger:00,hirschi:04}). The way this is
typically done is to make the assumption that the star has constant
angular velocity on spherical shells. In order to compute the
effective specific centrifugal force acting on a fluid parcel,
we compute the angular average of $(\vec{\omega} \times \vec{r})^2$
on a spherical shell of radius $r$, which leads to $f_\mathrm{cent} =
2/3\, \omega^2 r$. In Newtonian Lagrangian calculations, specific angular
momentum $j = \omega r^2$ is conserved by construction and the
effective centrifugal force appears in the momentum
equation. Relatively recently, such an approach has also been taken in
the Newtonian 1D core collapse calculations of
\cite{thompson:05,ott:06spin} in order to take into account the effect
of rotation approximately. In the Eulerian frame and in GR the
situation is more complicated.  We must solve an equation for angular
momentum conservation on top of taking into account a centrifugal
force term in the momentum equation. We begin by defining an azimuthal
Eulerian velocity $v^\phi$($=\omega$) and, in order to obtain a
quantity of dimension velocity, we also define $v_\varphi = r v^\phi$
(note that $u^\phi = W v_\varphi / r$).  With finite $v^\phi$,
$T^{r\phi}$ is finite and $W$ becomes $W = (1 - v^2 - 2/3
v_\varphi^2)^{-1/2}$ in our effective approach. We provide derivation
details in \ref{sec:appevolution} and present here only the
results. The modified stress-energy tensor leads to an additional
equation for angular momentum conservation analogous to
\eref{eq:evolution},
\begin{equation}
\partial_t(S_\phi) + {1
  \over r^2}\partial_r\left({\alpha r^2 \over X} F_\phi\right) = {\mathcal{S}}_\phi\,,
\end{equation}
where
\begin{eqnarray}
  \nonumber
  S_\phi &=& \rho h W^2 v_\varphi r\,\,,\\
  \nonumber
  F_\phi &=& \rho h W^2 v_\varphi r v = S_\phi v\,\,,\\
  {\mathcal{S}}_\phi &=& { \rho h W^2 \alpha v v_\varphi X} \left[ 
    4\pi r^2 P  +{m \over r}\right]\,\,.
\label{eq:conservedvars}
\end{eqnarray}
Also, an additional term, accounting for the centrifugal force, 
\begin{equation}
+ \alpha \frac{2}{3} \left({\rho h W^2 {v_\varphi}^2 \over Xr}\right)\,\,,
\end{equation}
appears on the RHS of the equation for $S^r$. Finally, the change
of the stress-energy tensor also has an effect on the metric
potential $\Phi$, whose equation is now given by
\begin{equation}
\partial_r \Phi = X^2 \left[\frac{m}{r^2} + 4\pi
  r\left(\rho h W^2 (v^2 + \frac{2}{3} v_\varphi^2) + P + \tau^\nu_\Phi\right)
\right]\,\,.
\label{eq:phirotation}
\end{equation}
We implement this 1.5D treatment of rotation in \code{GR1D}, but keep
the metric diagonal. The 1.5D treatment should be rather accurate for
slow rotation, and, as shown by \cite{ott:06spin}, will still capture
qualitatively the effect of centrifugal support due to rapid rotation.
For completeness, we note that the total angular momentum of the
system (see, e.g., \cite{cook:92}) is given by,
\begin{equation}
J = \int_0^\infty T^t_\phi \sqrt{-g}\, d^3x = {8 \pi \over 3} \int_0^\infty \rho h X W^2 r v_\varphi r^2dr\,\,,
\label{eq:totalJ}
\end{equation}
where we include a factor of 2/3 to account for the angular average.
The rotation parameter $\beta$, defined as the ratio
$T/|W_{\mathrm{grav}}|$ of rotational kinetic to gravitational energy
is
\begin{equation}
T/|W_{\mathrm{grav}}| = \frac{T}{|M_\mathrm{grav} - M_\mathrm{proper} - T|}\,\,,
\end{equation}
where
\begin{equation}
T = \frac{1}{2} \int_0^\infty \omega T^t_\phi \sqrt{-g} d^3 x = 
{4 \pi \over 3} \int_0^\infty \rho h X W^2 v^2_\varphi r^2 dr\,\,,
\end{equation}
where again a factor of 2/3 in the last step is from performing an
angular average.  $M_\mathrm{proper}$ is given by,
\begin{equation}
M_\mathrm{proper} = 4\pi \int_0^\infty (\rho + \rho\epsilon)\, X W  r^2 dr\,\,,
\end{equation}
and $M_\mathrm{grav}$ is specified by \eref{eq:mgrav}.

\section{Equations of State (EOS)}
\label{sec:EOS}

An EOS is needed to close the system of GR hydrodynamics equations and
provide the pressure as well as other thermodynamic quantities as a
function of density, temperature (or specific internal energy), and
composition. In \code{GR1D}, we include for test simulations the
standard analytic polytropic (isentropic ``cold'', $P=K\rho^\Gamma$)
and the $\Gamma$-law EOS (``hot'', $P=(\Gamma-1)\rho\epsilon$). These
are inappropriate for stellar collapse since they do not capture the
stiffening of the EOS at nuclear density. An analytic EOS, able to
capture this effect qualitatively and include nonisentropic effects,
is the \emph{hybrid} EOS \cite{janka:93} which we include in
\code{GR1D} and discuss in \sref{sec:hybrideos}. For a more realistic
description of the thermodynamics of nuclear matter, an EOS built from
a microphysical finite-temperature model for nuclear matter is
needed. This is also a prerequisite for any kind of neutrino
treatment, since crucial compositional information as well as chemical
potentials must be derived from a microphysical model. Such
microphysical EOS are too complicated to be computed on the fly in a
simulation and are used in tabulated form with
interpolation. \code{GR1D} is able to handle such EOS and we provide
tables at \code{http://www.stellarcollapse.org/microphysics} for the
EOS of Lattimer \& Swesty (\cite{lseos:91}, LS EOS) and for the one of
H.~Shen~et~al.~(\cite{shen:98b,shen:98a}, HShen EOS). The details of
these tables and the routines facilitating their use are discussed in
sections~\ref{sec:lseos} and \ref{sec:hsheneos}.

\subsection{Hybrid EOS}
\label{sec:hybrideos}

The hybrid EOS found widespread use in early multi-D simulations of
rotating core collapse (e.g., \cite{zwerger:97,dimmelmeier:02}), but
was shown by \cite{dimmelmeier:07,dimmelmeier:08} to lead in some
cases to qualitatively incorrect results for the collapse dynamics and
the resulting gravitational wave signal. We include it in \code{GR1D},
because its analytic nature provides for very fast calculations,
allowing us to readily test the GR hydrodynamics of \code{GR1D}. 

The hybrid EOS splits the pressure into a polytropic (cold) and a
thermal component,
\begin{equation}
P = P_\mathrm{cold} + P_\mathrm{thermal}\,\,.
\end{equation}
The cold part is piecewise polytropic. It is composed of a polytropic
EOS with $\Gamma = \Gamma_1$ for densities below nuclear
($\rho_\mathrm{nuc}$) and another polytropic EOS with $\Gamma =
\Gamma_2$ for densities above $\rho_\mathrm{nuc}$. The two are
smoothly matched at $\rho_\mathrm{nuc}$ which makes the polytropic
constant $K_2$ of the high-density part a function of the two
$\Gamma$s, of $K_1$, and of the transition density $\rho_\mathrm{nuc}$
(see, e.g. \cite{janka:93,zwerger:97,read:09a} for a description of
the procedure and detailed expressions). The thermal part is modeled
via a $\Gamma$-law with $\Gamma_\mathrm{th}$. It becomes relevant only
after core bounce when shocks are present, making the flow
nonadiabatic.  Its contribution is determined via the thermal specific
internal energy which is the difference between the primitive
variable $\epsilon$ and the cold specific internal energy,
$\epsilon_\mathrm{th} = \epsilon - \epsilon_\mathrm{cold}$.

For collapse simulations, we set $K_1 =
1.2435\times10^{15}(Y_e)^{4/3}\,\mathrm{[cgs]}$ (the value appropriate
for a relativistic degenerate gas of electrons,
\cite{zwerger:97,shapteu:83}) with $Y_e=0.5$.  We choose a value below,
but close to $4/3$ for $\Gamma_1$ and typically set $\Gamma_2 = 2.5$
to mimic the stiff nuclear EOS above $\rho_\mathrm{nuc}$ which we set
to $2\times10^{14}\,\mathrm{g\,cm}^{-3}$.  $\Gamma_\mathrm{th}$ we
normally keep at $1.5$ to model a mixture of relativistic ($\Gamma =
4/3$) and nonrelativistic ($\Gamma = 5/3$) thermal contributions. This
leads to rapid shock propagation and explosion. When simulating BH
formation with the hybrid EOS, we set $\Gamma_\mathrm{th}$ to smaller
values. This reduces the postshock thermal pressure and leads to shock
stagnation.

\subsection{Lattimer-Swesty EOS}
\label{sec:lseos}

The LS EOS \cite{lseos:91} is derived from a finite-temperature
compressible liquid-droplet model \cite{lattimer:85} with a Skyrme
nuclear force, uses the single heavy nucleus approximation, and
assumes nuclear statistical equilibrium (NSE).  NSE holds at $T
\gtrsim 0.5\,\mathrm{MeV}$ which in core collapse and supernova matter
is typically the case at $\rho \gtrsim \mathrm{few}\,\times 10^7
\,\mathrm{g\,cm}^{-3}$.

The LS EOS routines are open source and available from the Stony Brook
group\footnote{{\tt
    http://www.astro.sunysb.edu/dswesty/lseos.html}}. We employ their
baryonic parts to generate tables with nuclear incompressibilities
$K_0$ of $180\mev$, $220\mev$, and $375\mev$ (the larger $K_0$, the
stiffer the nuclear EOS). Hereafter, we refer to these $K_0$-variants
of the LS EOS as LS180, LS220, and LS375.  The symmetry energy $S_v$
is set in all variants to $ 29.3\mev $ for all $K_0$. Electrons and
photons are added using the routines provided by Timmes's
EOS\footnote{{\tt
    http://cococubed.asu.edu/code\_pages/eos.shtml}}~\cite{timmes:99}.

We compute the maximum cold neutron star masses for the three LS EOS
variants by setting $T=0.1\mev$ and assuming neutrino-less
$\beta$-equilibrium. The results are $1.83 \,M_\odot$ ($2.13 \,
M_\odot$), $2.04 \,M_\odot$ ($2.41 \, M_\odot$), $2.72 \,M_\odot$
($3.35 \, M_\odot$) for gravitational (baryonic) mass and for $K_0 =
180\mev$, $K_0 = 220\mev$, and $K_0 = 375\mev$, respectively.  The
coordinate radii of these maximum mass stars are
$10.1$~km, $10.6$~km, and $12.3$~km.

Our LS EOS tables have 18 evenly-spaced points per decade in
$\log_\mathrm{10} \rho$ ranging from $10^3 - 10^{16}\, \mathrm{g\,
  cm}^{-3}$, 30 points per decade in $\log_\mathrm{10} T$ ranging from
$10^{-2} - 10^{2.4}\mev$, and 50 points equally spaced in electron
fraction from 0.035 to 0.53.  This table resolution is sufficiently
good to allow the use of simple and fast tri-linear interpolation (in
$\log_{\mathrm{10}}(\rho),\ \log_{\mathrm{10}}(T),\ Y_e$),  in
collapse simulations while maintaining good thermodynamic
consistency. In tests of adiabatic collapse, the inner-core entropy is
conserved to $\sim 1\%$ from the onset of collapse to core bounce.

To generate the LS EOS tables, we employ the LS EOS at densities above
$10^8\,\mathrm{g\,cm}^{-3}$, but, due to unreliable convergence, use
linear extrapolation of the Helmholtz free energy $F$ in $Y_e$ for
$Y_e > 0.5$ and in $T$ at $T < 0.06 \mev$. Note that the latter is far
away from NSE, but is never reached by core collapse trajectories at
$\rho > 10^8\,\mathrm{g\,cm}^{-3}$. At densities below
$10^8\,\mathrm{g\,cm}^{-3}$, we use the Timmes EOS~\cite{timmes:99}
and assume that the matter is an ideal gas composed of electrons,
photons, neutrons, protons, alpha particles, and heavy nuclei with the
average $A$ and $Z$ given by the LS EOS at the transition.  

Since the specific internal energies returned by the baryonic part of
the Timmes EOS do not contain the nuclear binding energy, we shift the
zero point of the Timmes EOS so that the returned specific internal
energies are consistent with the LS EOS values at the transition
point.  For simplicity, we keep baryonic compositional variables fixed
at the values obtained from the LS EOS at the transition density.
These particular choices for the baryonic component have little effect
at low densities where the thermodynamics are dominated by electrons
at low to intermediate temperatures and by photons at high
temperatures.  However, for full core-collapse supernova simulations
that intend to address also nuclear burning and nucleosynthesis
aspects, a more involved consistent NSE/non-NSE EOS treatment
involving the advection of many chemical species and a treatment of
their interactions with a nuclear reaction network is necessary. We
will leave such a treatment to future work (but see, e.g.,
\cite{rampp:02,buras:06a} for discussions of such implementations).

When using finite-temperature microphysical NSE EOS such as the LS EOS
in GR hydrodynamics codes, two additional caveats need to be taken
into account: \emph{(1)} The thermodynamic potential from which all
dependent variables are derived is the Helmholtz free energy $F$. This
makes the EOS a function of $\{\rho,T,Y_e\}$ while GR hydrodynamics
codes such as \code{GR1D} operate on the primitive thermodynamic and
compositional variables $\{\rho,\epsilon,Y_e\}$. Hence, in a typical
EOS call it is first necessary to determine $T(\rho,\epsilon,Y_e)$
through a root-finding procedure, before the dependent variables can
be obtained through tri-linear interpolation in
$\{\rho,T,Y_e\}$. \emph{(2)} In contrast to Newtonian hydrodynamics
that involves only differences of the specific internal energy
$\epsilon$, GR codes depend directly on $\epsilon$ through its
contribution to the matter stress-energy tensor. Hence, it is
important to find and use a physically correct energy zero point and
ensure that there are no rest-mass contributions included in
$\epsilon$.

\subsection{HShen EOS}
\label{sec:hsheneos}
The HShen EOS \cite{shen:98a,shen:98b} is based on a relativistic
mean-field model for nuclear interactions, assumes NSE, and is
extended with the Thomas-Fermi approximation to describe the
homogeneous phase of matter as well as the inhomogeneous matter
composition.  $K_0$ of the HShen EOS is $ 281\mev $ and the symmetry
energy $S_v$ has a value of $ 36.9\mev $. The authors of the HShen EOS
provide the baryonic component\footnote{{\tt
    http://user.numazu-ct.ac.jp/$\sim$sumi/eos}} in tabulated form
only. The provided table is not uniformly spaced and has too low
resolution to be used directly with fast tri-linear interpolation in
simulations. Hence, we generate a finer uniformly-spaced table that
has 18 points per decade in $\log_\mathrm{10} \rho$ from $10^3 -
10^{15.36}\,\mathrm{g\,cm}^{-3}$, 41 points per decade in
$\log_\mathrm{10} T$ from $10^{-2} - 10^{2.4}\mev$, and 50 points in
$Y_e$ covering the interval $0.015 - 0.56$.  We interpolate all
dependent variables from the original HShen table using the cubic
Hermite interpolation function given in \cite{timmes:00} modified to
have monotonic interpolation behavior according to the prescription of
\cite{steffen:90}. The interpolation is performed first bicubic in
${\rho,T}$, then cubic in $Y_e$. Alternatively to the just described,
one could interpolate the Helmholtz free energy $F$ and re-derive
dependent variables by taking derivatives of $F$ on the interpolated
table (see, e.g., \cite{timmes:00}). We decided against this approach,
since it would require quintic interpolation and the knowledge of the
second derivatives of $F$ at each point in the original table,
some of which would have to be computed by taking second derivatives
in the coarse original table. Also, compositional information cannot
be obtained directly from $F$ and would have to be interpolated
from the original table.

We perform the described interpolation at densities above
$10^{7}\,\mathrm{g\,cm}^{-3}$. For points with $T > 100\mev$ and $T <
0.1\mev$ we extrapolate most variables linearly, keeping only the
compositions fixed. We add photons and electrons after interpolation
using the routines of the Timmes EOS.  At densities below
$10^{7}\,\mathrm{g\,cm}^{-3}$, we employ the Timmes EOS in the same
fashion as described in the above for the LS EOS.

We compute the maximum cold neutron star masses for the HShen EOS in
the same way as for the LS EOS and find $2.24 \,M_\odot$ and $2.61 \,
M_\odot$, for the gravitational and baryonic value, respectively. The
coordinate radius of the corresponding star is $12.6$~km.

\section{Neutrino Leakage and Heating}
\label{sec:leakage}

\subsection{Deleptonization and Electron Capture 
in the Collapse Phase}
\label{sec:fakenu}

Electron capture on free and bound protons leads to the emission of
neutrinos that stream away from the core and carry away net lepton
number at densities below $\sim 10^{12}\,\mathrm{g\,cm^{-3}}$. Hence,
one speaks of the deleptonization of the core.
The change of the electron fraction $Y_e$ in the collapse phase due to
deleptonization has important dynamical consequences.  A
reduction of $Y_e$ leads to a decrease of the mass of the homologously
collapsing inner core whose kinetic energy is initially imparted on
the supernova shock and which turns into the PNS core after bounce
\cite{bethe:90}. We take electron capture in collapse into account in
\code{GR1D} with the approximate scheme of Liebend\"orfer
\cite{liebendoerfer:05fakenu} who observed that $Y_e$ of infalling
mass elements depends primarily on the local matter density
$\rho$ and can be parameterized with rather high precision on the basis
of radiation-hydrodynamic calculations.

\noindent Operator-split, after a hydrodynamics update, we compute
the change in $Y_e$,
\begin{equation}
\Delta Y_e = \min \left[0, \overline{Y}_e(\rho) - Y_e \right]\,\,,
\end{equation}
which ensures for consistency that a change in $Y_e$ is either
negative or $0$. We use for $\overline{Y}_e(\rho)$ the fitting formula
given in \cite{liebendoerfer:05fakenu} with parameters $\rho_1 =
3\times10^7\,{\rm{g\ cm^{-3}}} $, $\rho_2 = 2\times10^{13}\,{\rm{g\ {cm}^{-3}}}$,
$Y_1 = 0.5$, $Y_2 = 0.278$, and $Y_c = 0.035$ corresponding to the
15-$M_\odot$ model of \cite{ww:95}, evolved as model G15 by
\cite{liebendoerfer:05}. \code{GR1D} also contains an interpolation
routine to use numerical $\overline{Y}_e(\rho)$ data.

Electron capture leads to a change in the entropy ($s$, the specific
entropy in units of $k_\mathrm{B} / \mathrm{baryon}$) that is carried
away by neutrinos leaving the core at densities below an assumed
trapping density $\rho_\mathrm{trap} = 2 \times
10^{12}\,\mathrm{g\,cm}^{-3}$. The entropy change is 
given by
\begin{equation}
\Delta s = -\Delta Y_e \frac{\mu_p - \mu_n + \mu_e - E_\nu}{k_\mathrm{B} T}\,\,.
\end{equation}
$E_\nu$ is the energy of the escaping neutrinos (set to
$10\mev$). $\mu_p$, $\mu_n$, and $\mu_e$ are the proton, neutron, and
electron chemical potentials including rest mass, respectively. Following
\cite{liebendoerfer:05fakenu}, we set $\Delta s = 0$ if $\mu_p + \mu_n
+ \mu_e < E_\nu$ and above $\rho_\mathrm{trap}$. After updating the
entropy, we use the EOS to update the specific internal energy $\epsilon$
for consistency with the new $Y_e$ and $s$.

We employ the outlined deleptonization scheme until core bounce
(defined as the time when the peak entropy of the inner core surpasses
3 $k_B$/baryon) and until 5ms after bounce for yet unshocked regions
of the outer core that will settle in the high-density outer PNS and
only in this way assume realistic postbounce Ye.

\subsection{Postbounce Deleptonization and Neutrino Heating/Cooling}

At core bounce a strong hydrodynamic shock wave is generated that
travels outward into the outer core, heating and dissociating
infalling heavy nuclei into nucleons. Electron capture occurs rapidly
on free protons and a sea of electron neutrinos ($\nu_e$) builds up
and is released in the $\nu_e$ burst when the shock breaks through the
neutrinosphere\footnote{The neutrinosphere is the
  effective ``decoupling'' surface of neutrinos where the optical depth
  $\tau_\nu$ of the supernova matter is $2/3$. Its position depends
  strongly on neutrino energy.}, deleptonizing the postshock region
and leaving behind a ``trough'' in the $Y_e$ profile (e.g.,
\cite{thompson:03}). The softening of the EOS due to dissociation of
nuclei and postshock energy loss to escaping neutrinos lead the shock
to stall and turn into an accretion shock soon after bounce. In the
hot postshock region, electrons are less degenerate and positrons
appear and are captured on neutrons, leading to a rise of the
$\bar{\nu}_e$ luminosity.  In addition, in the PNS and in the
postshock region, neutrinos and antineutrinos of all flavors are
emitted by thermal processes.

The simple $\overline{Y}_e(\rho)$ parameterization discussed in the
previous section~\ref{sec:fakenu} is not adequate to capture these
effects and, in principle, a full neutrino energy-dependent
radiation-hydrodynamics treatment would be needed for accurately
capturing postbounce neutrino effects.  Such a treatment may be
added in future versions of \code{GR1D}.  In the present version of
\code{GR1D}, we approximate postbounce neutrino transport by a gray
(energy-averaged) neutrino leakage scheme augmented with a simple
prescription for neutrino heating in the postshock region.  This
approach captures the most important qualitative aspects of the
postbounce evolution well and, as we demonstrate in
\sref{sec:s40_leakage}, is sufficiently quantitatively accurate 
to make reliable predictions of the time of BH formation and the
maximum PNS mass in failing core-collapse supernovae.

Our implementation in \code{GR1D} combines elements of the neutrino
leakage schemes of Ruffert~et~al.~\cite{ruffert:96} and of Rosswog \&
Liebend\"orfer \cite{rosswog:03b}. We consider three neutrino species,
$\nu_{e}$, $\bar{\nu}_{e}$, and $\nu_{x}$. In the latter, we lump
together $\mu$ and $\tau$ neutrinos and anti-neutrinos since they
interact only by neutral-current processes in the core collapse
context and have very similar cross sections. 
The mean (energy-averaged) optical depth is 
\begin{equation}
\tau_{\nu_i}(r) = \int_{r}^{\infty} \kappa_t(\nu_i)\, X dr\,\,,
\label{eq:tau}
\end{equation}
where $\kappa_t(\nu_i)$ is the mean transport opacity equal to the sum
of absorptive and scattering opacities\footnote{Note that the
  opacities for neutrino number and neutrino energy transport
  differ. Hence, the optical depths for number and energy transport
  must be computed separately \cite{ruffert:96}. We neglect this
  subtlety and use the optical depths for energy transport throughout
  \code{GR1D}.}  for neutrino species $\nu_i$. We follow
\cite{ruffert:96} in the calculation of $\kappa_t(\nu_i)$ and of the
approximate neutrino degeneracy parameters ($\eta_{\nu_i} =
\mu_{\nu_i}/T$). We consider opacity contributions from neutrino
scattering on neutrons, protons, and heavy nuclei and absorption of
neutrinos (anti-neutrinos) on neutrons (protons). For heavy-lepton
neutrinos that are never degenerate, we set $\eta_{\nu_x} =
0$. $\eta_{\nu_e}$ is known {(\emph{1})} in $\beta$-equilibrium where
$\eta_{\nu_e}^\mathrm{eq} = \eta_e + \eta_p - \eta_n$ (where we assume
that the chemical potentials include rest mass terms) and {(\emph{2})}
in the free streaming limit, where $\eta_{\nu_e}^\mathrm{stream} =
0$. Furthermore, $\eta^{\rm{eq}}_{\bar{\nu}_e} = - \eta^{\rm{eq}}_{\nu_e}$.
In between the two regimes, the neutrino distribution
function cannot be derived from first principles and neutrino
transport is necessary for a correct estimate of $\eta_{\nu_e}$ and $\eta_{\bar{\nu}_e}$. As an
approximation, we interpolate between {(\emph{1})} and {(\emph{2})}
using the optical depth,
\begin{equation} 
\eta_{\nu_i} = \eta_{\nu_i}^\mathrm{eq} (1 - e^{-\tau_{\nu_i}(\eta_{\nu_i})})\,\,.
\label{eq:eta}
\end{equation}

\noindent 
Note that $\tau_{\nu_i}$ depends on $\eta_{\nu_i}$ and vice versa.
Hence, we iterate their calculation until convergence is
reached\footnote{Initially we choose $\kappa_{\nu_i}(r) =
  10^{-5}{\rm{cm}}^{-1}$ determine $\tau_{\nu_i}$ through \eref{eq:tau} and
  iterate \eref{eq:eta}.  For all subsequent times we use the
  previously determined value of $\tau_{\nu_i}$ as a starting point,
  convergence (fractional difference in $\kappa_{\nu_i} < 10^{-10}$)
  is typically reached after three iterations.}.

Knowing $\tau_{\nu_i}$ and $\eta_{\nu_{i}}$, we use the leakage scheme
of \cite{rosswog:03b} to calculate the neutrino emission rates for the
capture processes $p+e^- \to \nu_e + n$ and $e^+ + n \to \bar{\nu}_e +
p$ and thermal emission via electron-positron annihilation and plasmon
decay to $\nu\bar{\nu}$ pairs.  We modify the scheme of
\cite{rosswog:03b} in the following ways: (\emph{i}) we use the
interpolated $\eta_{\nu_i}$ from above instead of the equilibrium
values suggested in \cite{rosswog:03b}, (\emph{ii}) we increase their
diffusion time scale $t^\mathrm{diff}_{\nu_i}$ by a factor of 2 to
obtain more reasonable neutrino luminosity predictions, and
(\emph{iii}) for simplicity, we use the analytic thermal emissivities
from \cite{ruffert:96}. Following \cite{rosswog:03b}, we then
interpolate the effective volumetric energy loss
$Q_\mathrm{eff}^\mathrm{leak}$ (${\rm{erg}} / \rm{cm}^3 / \rm{s}$) and
effective number loss $R_\mathrm{eff}^\mathrm{leak}$ ($\# /
{\rm{cm}}^3 /{\rm{s}}$) between the limits of diffusive emission
(subscript ``diff'') and free emission (subscript ``loc'') using
\begin{equation}
\chi^\mathrm{leak}_{\mathrm{eff},\nu_i} = \chi^\mathrm{leak}_{\mathrm{loc},\nu_i} / 
( 1 + \chi^\mathrm{leak}_{\mathrm{loc},\nu_i}/\chi^\mathrm{leak}_{\mathrm{diff},\nu_i})\,\,,
\end{equation} 
where $\chi = Q$ for energy loss and $\chi=R$ for number loss (see
\cite{rosswog:03b} for definitions and details). We define the
neutrino luminosity seen by an observer at rest at radius $r$ in the
coordinate frame by summing up the effective energy emission rates
from each zone interior to $r$, transforming from the fluid rest frame
(FRF) to the coordinate frame (CF), and applying the redshift (see
\ref{sec:appendixlum} for details),
\begin{equation}
\fl L^{\rm{CF}}_{\nu_i} (r) = 4\pi \int_0^r \left[ {\alpha(r') \over \alpha(r)}\right]\,
Q_{\mathrm{eff},\nu_i}(r') [\alpha(r') W(r') (1 + v(r')) ] X(r') {r'}^2 dr'\,\,.\label{eq:nulumCF}
\end{equation}
For an observer at rest at $r=\infty$ $\left(\alpha(\infty) = 1\right)$,
\begin{equation}
\fl L_{\nu_i} (\infty) = 4\pi \int_0^\infty \alpha(r')\,
Q_{\mathrm{eff},\nu_i}(r') [\alpha(r') W(r') (1 + v(r')) ] X(r') {r'}^2 dr'\,\,.
\end{equation}
It is useful to note the neutrino luminosity as seen by an
observer at rest in the fluid rest frame at radius $r$,
\begin{equation}
L^{\rm{FRF}}_{\nu_i} (r) = \frac{L^{\rm{CF}}_{\nu_i}(r)}{\alpha(r) W(r)
(1 + v(r) )}\,\,,\label{eq:nulumFRF}
\end{equation}
where the denominator transforms the luminosity from the frame of an
observer at rest in the coordinate frame \eref{eq:nulumCF} to the
fluid rest frame. 

\subsubsection{Neutrino Heating.}
In addition to the above leakage scheme, we include a parameterized
heating scheme to mimick neutrino absorption in the postshock
region. Heating occurs at intermediate to low optical depths where
neutrinos begin to decouple from matter and a net energy transfer from
neutrinos to the fluid is possible~(see, e.g., \cite{janka:01}).  The
dominant heating processes are the charged-current capture reactions
of $\nu_e$ on neutrons and $\bar{\nu}_e$ on protons.  We take the
absorption cross sections from \cite{rosswog:03b},
\begin{eqnarray}
\sigma_{\mathrm{heat},\nu_e} &=& {(1 + 3 g_A^2) \over 4} \sigma_0 {
 \langle\epsilon^2\rangle_{\nu_e}^\mathrm{ns} \over (m_ec^2)^2}\langle 1 -
f_{e^-}\rangle\,\,, \\ [0.2 mm]
\sigma_{\mathrm{heat},\bar{\nu}_e} &=& {(1 + 3 g_A^2) \over 4} \sigma_0 {
 \langle\epsilon^2\rangle_{\bar{\nu}_e}^\mathrm{ns} \over (m_ec^2)^2}\langle 1 -
f_{e^+}\rangle\,\,,
\end{eqnarray}

\noindent
where $\sigma_0$ is a reference weak-interaction cross section equal
to $1.76\times10^{-44}\,\rm{cm}^2$, $g_A \sim -1.25$, and the Fermi
blocking factors $\langle 1 - f_{i}\rangle$ are defined analogously to
\cite{ruffert:96,rosswog:03b}.  In the postshock region the positron
blocking term is negligible but the electron blocking term can be
significant around the time of bounce.  Following
\cite{janka:01}, we set the mean squared neutrino energy to
$\langle\epsilon^2\rangle_{\nu_i}^\mathrm{ns} = T(\tau_{\nu_i} =
\frac{2}{3})^2 {\cal{F}}_5(\eta_{\nu_i}^\mathrm{ns}) /
{\cal{F}}_3(\eta_{\nu_i}^\mathrm{ns})$, where $T(\tau_{\nu_i}=\frac{2}{3})$
is the temperature at the neutrinosphere of species $i$, superscript $\mathrm{ns}$ denotes neutrinospheric values, and
${\cal{F}}_n(\eta) = \int_0^\infty {x^n dx \over \exp(x-\eta)+1}$ is
the $n^{\rm{th}}$ Fermi integral (we approximate Fermi integrals via
the formulae given in \cite{takahashi:78}).

Given the neutrino luminosity $L^{\rm{FRF}}_{\nu_i} (r)$ obtained from the
leakage scheme \eref{eq:nulumFRF}, we write the local neutrino
heating rate in units of $\mathrm{erg\, cm}^{-3}\,\mathrm{s}^{-1}$ as
\begin{equation}
Q^{\mathrm{heat}}_{\nu_i}(r) = f_\mathrm{heat} \frac{L^{\rm{FRF}}_{\nu_i}(r)}{4\pi r^2}
\sigma_{\mathrm{heat},\nu_i}\, {\rho\over m_u} X_i \left\langle {1 \over
  F_{\nu_i}} \right\rangle e^{-2\tau_{\nu_i}} \,\,,\label{eq:Qheat}
\end{equation}
where $m_u$ is atomic mass unit and the mass fraction $X_i = X_n$ in
the case of $\nu_e$ absorption and $X_i = X_p$ for
$\bar{\nu}_e$s. $\langle 1/F_{\nu_i} \rangle$ is the mean inverse flux
factor describing the degree of forward-peaking of the radiation field
(e.g., \cite{ott:08,janka:01}; $\langle 1/F_{\nu_i} \rangle$ is 1 for
free streaming and diverges at high optical depth). We estimate
$\langle 1/F_{\nu_i} \rangle$ by the interpolation $\langle
1/F_{\nu_i}(\tau) \rangle = 4.275\tau + 1.15$, which reproduces the
predicted values of $4$ at the neutrinosphere \cite{janka:01} and
levels off at a value of $1.15$ at low optical depth in the outer
postshock region. We choose the latter value instead of $1$, because
(\emph{a}) the radiation field becomes fully forward peaked only
outside the shock (e.g., \cite{ott:08}), and (\emph{b}) the linear
interpolation in $\tau$ drops off too quickly compared to full
simulations \cite{ott:08}, hence the higher floor value to
compensate. Finally, we introduce the attenuation factor
$e^{-2\tau_{\nu_i}}$ to cut off heating near and below the
neutrinosphere and the scaling factor $f_\mathrm{heat}$ to allow for
an ad-hoc increase of the heating rate.  Once the heating rate for a
computational cell is computed, we reduce the outgoing luminosity by
the deposited power for overall energy conservation. In the coordinate
frame \eref{eq:nulumCF} now becomes,
\begin{equation}
\fl L^{\rm{CF}}_{\nu_i} (r) = 4\pi \int_0^r \left[{\alpha(r') \over \alpha(r)}\right]
\left[Q_{\mathrm{eff},\nu_i}(r') -
  Q^{\mathrm{heat}}_{\nu_i}(r')\right] [\alpha(r') W(r') (1 + v(r'))] X(r') r'^2 dr'\,\,.
\label{eq:nulumCF_heat}
\end{equation}

Along with the energy deposition goes a change in $Y_e$
which can be written as
\begin{equation}
R^\mathrm{heat}_{Y_e} = \frac{Q_{\nu_e}^{\mathrm{heat}}}{\langle
  \epsilon_{\nu_e}^\mathrm{ns} \rangle} -
\frac{Q^{\mathrm{heat}}_{\bar{\nu}_e}}{\langle \epsilon_{\bar{\nu}_e}^\mathrm{ns}
  \rangle} \,\,,
\end{equation}
where we approximate the mean neutrino energies based on their
neutrinospheric values as $\langle \epsilon_{\nu_i}^\mathrm{ns} \rangle =
T(\tau_{\nu_e} =\frac{2}{3})
\mathcal{F}_5(\eta_{\nu_i}^\mathrm{ns})/\mathcal{F}_4(\eta_{\nu_i}^\mathrm{ns})$
\cite{rosswog:03b}.  

To caution the reader, we point out that the simple gray heating
scheme presented in the above is not self-consistent and cannot
replace a radiation transport treatment that allows emission and
absorption to balance. While we find that the combination of gray
leakage/heating reproduces the overall qualitative dynamical features
observed in postbounce radiation-hydrodynamic simulations,
quantitative aspects are not captured as well. This is true in
particular in highly dynamical situations shortly after bounce when
we observe an unphysical rise of the electron fraction due to heating
in the lower postshock region.

 We couple the neutrino leakage/heating scheme with
the GR hydrodynamics in \code{GR1D} through source/sink terms on the
RHS of the GR hydrodynamics equations in MoL.
Neutrino--matter interactions occur in the fluid rest frame where the
total energy and number changes are given by
\begin{equation}
\hspace*{-2cm}Q^0_\mathrm{E} = Q^\mathrm{heat}_\mathrm{total} - Q^\mathrm{leak}_\mathrm{eff, total}\,\,,\hspace*{0.4cm}
R^0_{Y_e} = R^\mathrm{heat}_\mathrm{total} + R^\mathrm{leak}_\mathrm{eff,total}\,\,,\label{eq:leakagerates}
\end{equation}
where $Q^\mathrm{heat}_\mathrm{total}$ and
$Q^\mathrm{leak}_\mathrm{eff, total}$ are always positive or zero and
$R^\mathrm{heat}_\mathrm{total}$ and
$R^\mathrm{leak}_\mathrm{eff,total}$ may be positive or negative.
Following \cite{pons:97,mueller:09phd}, transforming these terms to the
coordinate frame via the methods laid out in \ref{sec:appendix}, we
obtain the neutrino heating/cooling and deleptonization source/sink
terms for the RHS in the MoL integration,
\begin{equation}
\hspace*{-2cm}R_{Y_e}^\nu  = \alpha X R^0_{Y_e}\,\,,\hspace*{0.3cm}
Q_{S^r}^{\nu, E} = \alpha v W Q^0_\mathrm{E}\,\,,\hspace*{0.3cm}
Q_\tau^{\nu,E} = \alpha W Q^0_\mathrm{E}\,\,.
\label{eq:neutrinosources}
\end{equation}

\subsection{Neutrino Pressure}
\label{sec:neutrinopressure}
Electron neutrinos above trapping density in the inner core during
the final phases of collapse and in the postbounce PNS contribute to
both the pressure and the specific energy density (with relative importance
of up to $\sim 10\%$ around core bounce \cite{ott:06phd}).  We neglect
neutrino contributions to pressure and energy below
$\rho_\mathrm{trap}$ where they are small, but otherwise follow
\cite{liebendoerfer:05fakenu} and assume electron neutrinos and
antineutrinos to be a perfect Fermi gas. The pressure is then given by
\begin{equation}
P_\nu = {4 \pi \over 3 (hc)^3}T^4\left [F_3(\eta_\nu)
  +F_3(-\eta_\nu)\right]\,\,,
\end{equation}

\noindent
where $\eta_\nu=\mu_\nu /T$ and $\mu_\nu = \mu_e-\mu_n + \mu_p$, where
the chemical potentials include rest mass contributions.  $F_3$ is the
$3^\mathrm{rd}$ Fermi integral which we approximate following
\cite{bludman:77}.  The specific internal energy of a relativistic
Fermi gas of neutrinos is simply $\epsilon_\nu = {3 P_\nu / \rho}$.

We treat neutrinos and fluid separately from each other and treat
momentum transfer between the neutrino radiation field and the fluid
approximately using the radial gradient of the neutrino pressure as
suggested by \cite{liebendoerfer:05fakenu}.  We couple this radiation
stress into \code{GR1D}'s MoL integration of the GR momentum ($S^r$)
and energy ($\tau$) equations via source terms (see \ref{sec:appendix}
for a derivation; we neglect rotational effects in these source terms),
\begin{equation}
Q_{S^r}^{\nu, \mathrm{M}} = - \alpha W \frac{\partial P_\nu}{\partial r}
\,\,, \hspace*{2cm} Q_\tau^{\nu, \mathrm{M}} = - \alpha W v
\frac{\partial P_\nu}{\partial r}\,\,.
\label{eq:neutrinomatter}
\end{equation}

In addition to the force on the fluid due to the neutrino pressure
gradient, we take into account the energy and ``pressure'' of the
neutrino radiation field by adding $P_\nu$ and $\epsilon_\nu$ 
through the terms $\tau^\nu_m$ and $\tau^\nu_\Phi$
in \eref{eq:mgrav} and Eqs.~\eref{eq:phi} and \eref{eq:phirotation}.
These contributions are derived by modifying the stress-energy tensor,
\begin{equation}
T^{\alpha \beta} = \rho \left[ 1 + (\epsilon + \epsilon_\nu) +\left(
  {P + P_\nu \over \rho}\right)\right] u^\alpha u^\beta + g^{\alpha
  \beta} (P+P_\nu)\,\,,
\end{equation}
$\tau^\nu_m$ and $\tau^\nu_\Phi$ are then given by \cite{noble:03phd}
\begin{equation}
  \tau^\nu_m = \rho W^2 (\epsilon_\nu + P_\nu/\rho) -P_\nu = (4 W^2 
  - 1)P_\nu\,\, ,
\end{equation}
\begin{equation}
\tau^\nu_\Phi = \rho W^2 v^2 (\epsilon_\nu + P_\nu/\rho) + P_\nu = (4
W^2 v^2 + 1)P_\nu\,\, . 
\label{eq:taunuphi}
\end{equation}
We note that if rotation is included, $v^2$ in \eref{eq:taunuphi}
is replaced with $v^2 + {2 \over 3}v^2_\varphi$.

\section{Code Tests}
\label{sec:tests}

In the following, we provide results from a set of standard and
stringent relativistic hydrodynamics code tests for which analytic
results exist.  These involve two planar shocktube problems in
\sref{sec:shocktube}, the spherical Sedov blast wave problem in
\sref{sec:sedov}, and Oppenheimer-Snyder collapse in \sref{sec:oppen}.
Finally, in \sref{sec:hybrid}, we present results from a collapse
simulation of a $n=3$ polytrope and demonstrate convergence of
the hydrodynamics scheme in \code{GR1D}. With this selection, we test a
broad range of aspects of potential problems to be addressed with
\code{GR1D}: special relativistic effects, geometrical effects, and
fully general-relativistic collapse dynamics.

\subsection{Relativistic Shocktube}
\label{sec:shocktube}

\begin{figure}[b]
\begin{center}
\hspace*{-0.15cm}\includegraphics[width=7.1cm,clip=]{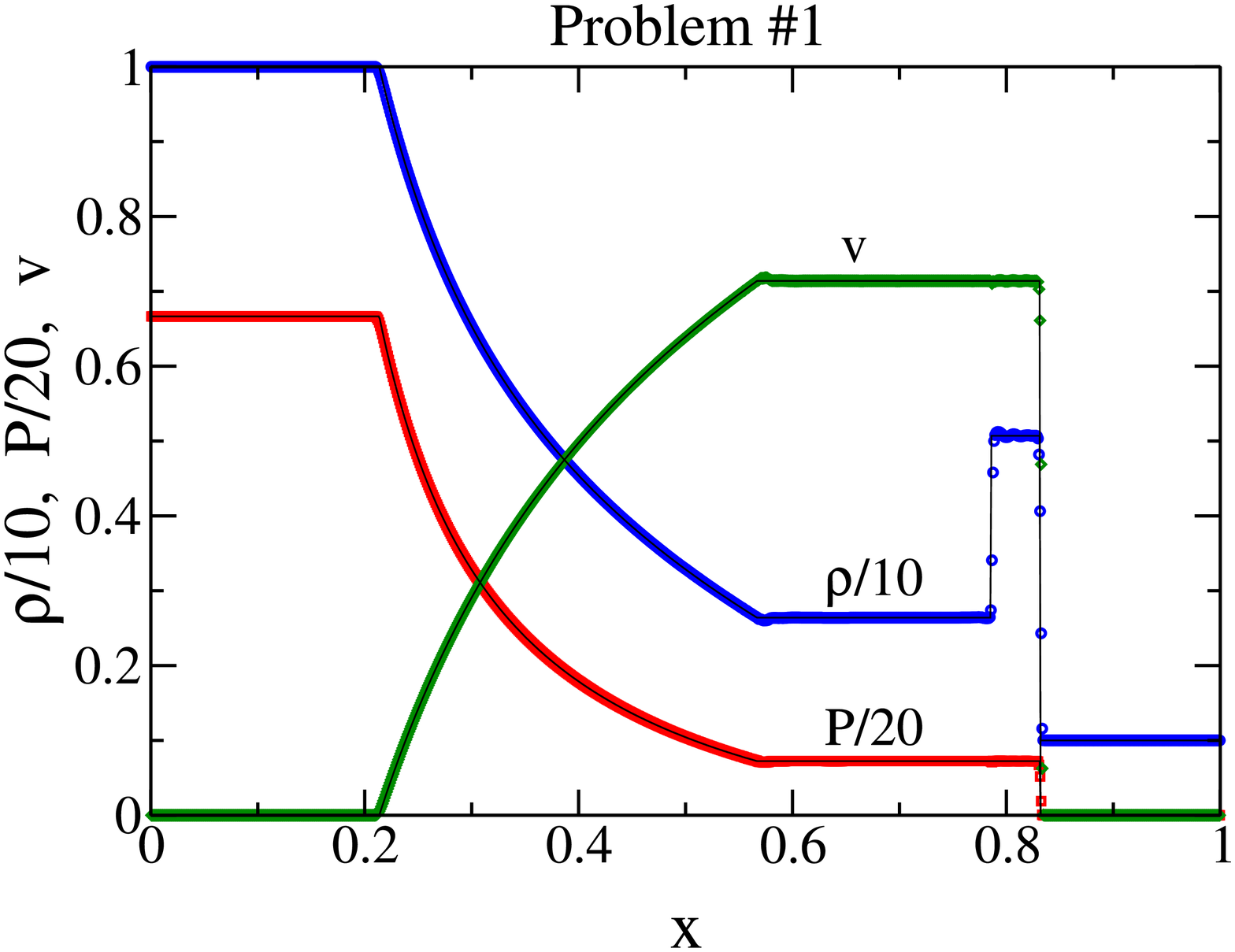} 
\hspace*{-0.7cm}\includegraphics[width=7.1cm,clip=]{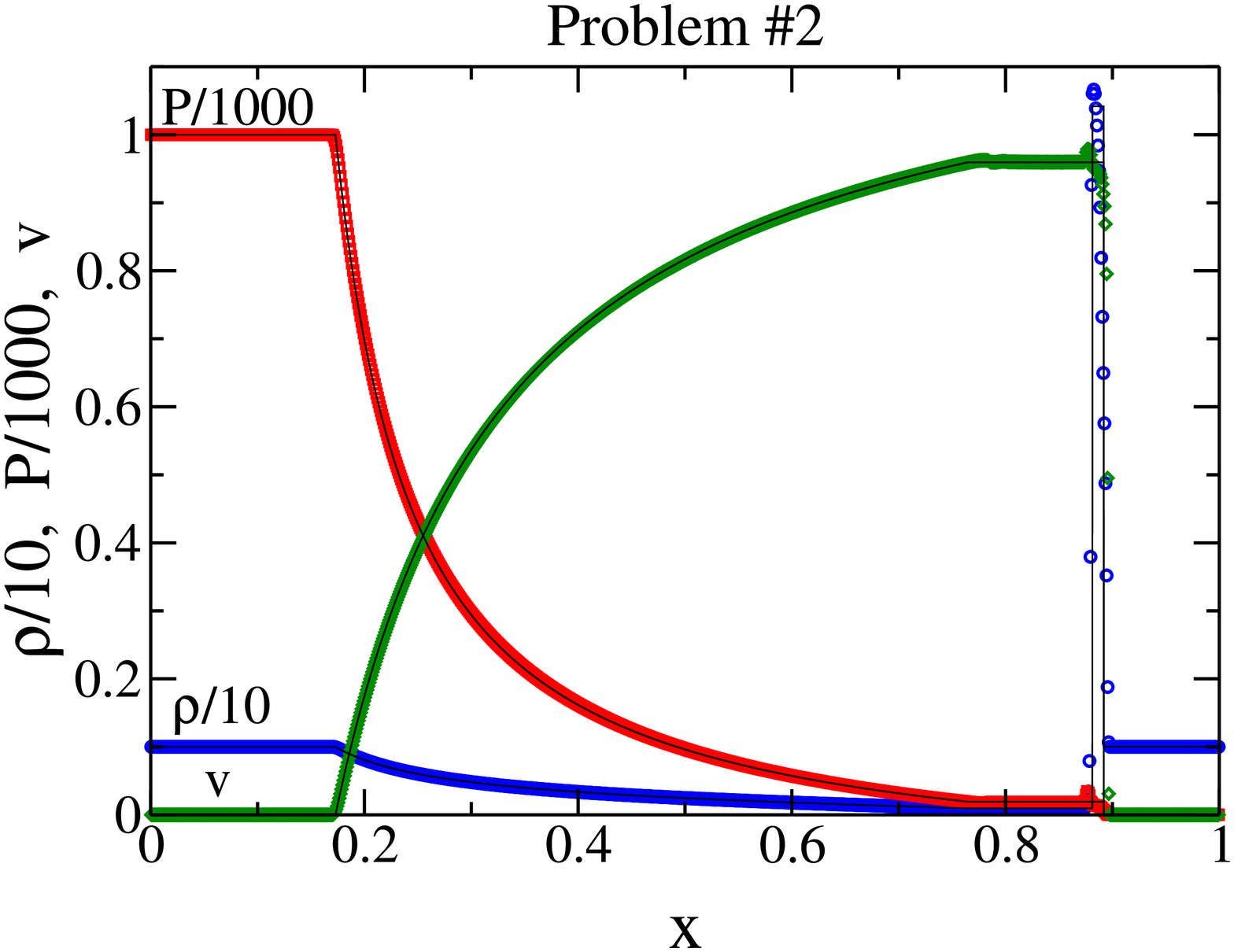}\hspace*{-0.4cm} 
\caption{Relativistic shocktube simulations: Initial conditions taken
  from \cite{marti:03} and provided in Table~\ref{tab:shocktube}. The
  pressure, density, and velocity are shown at $t=0.4$ for problem \#1
  (left panel) and problem \#2 (right panel).  For reference, in both 
  figures the
  pressure is denoted by boxes (red), density by circles (blue) and
  velocity by diamonds (green). The analytic solution is denoted by
  the solid line. Both problems were run with a Courant factor of 0.5
  and 3$^\mathrm{rd}$~order Runge-Kutta integration.}\label{fig:sodshockcomp}
\end{center}
\end{figure}

We assume flat space and planar geometry and perform the two
relativistic shocktube tests proposed by \cite{marti:03}. We use a
$\Gamma-$law EOS with $\Gamma = $~5/3 and a grid of length $1$ with a
cell spacing of ${{dx}} = 0.001$. The starting values of the density,
pressure and velocity are summarized in Table \ref{tab:shocktube}.
The left panel of \fref{fig:sodshockcomp} shows the exact results for
velocity, density, and pressure of the mildly-relativistic problem \#1
at $t=0.4$. Superposed are the numerical results obtained with
\code{GR1D} that reproduce the exact results nearly perfectly.
Problem \#2 is a more stringent test and involves Lorentz factors of
up to $6$ in the forward propagating shock and a very thin shell of
trailing matter. As shown in the right panel of
\sref{fig:sodshockcomp}, \code{GR1D} reproduces the exact solution at
$t=0.4$ very well almost everywhere, but fails to completely resolve
the thin shell of relativistic matter. This is most likely due to the
rather diffusive nature of the HLLE Riemann solver employed
in \code{GR1D} (see, e.g., \cite{dimmelmeier:02a,ott:06phd} for
comparable results obtained with a nominally more accurate scheme). In
an attempt to obtain results closer to the analytic solution we use
3$^{rd}$~order Runge-Kutta time integration for this test case.  These
deviations are not worrying since the shocks that obtain in
stellar collapse are much less relativistic than that of problem
\#2. If \code{GR1D} were to be applied to ultrarelativistic outflows
(e.g., in a GRB), a more precise treatment of the Riemann problem
would likely be necessary.

\begin{table}[t]
\centering
\begin{tabular}{c|c|c|c} \hline
\multicolumn{2}{c}{P1} & \multicolumn{2}{|c}{P2} \\ 
\hline \hline
$r < 0.5$ & $r>0.5$ & $r<0.5$ & $r > 0.5$ \\ 
\hline
$\rho = 10$ & $\rho = 1$ &  $\rho = 1$ &  $\rho = 1$ \\
$P = 13.33$ & $P = 0$ &  $P = 10^3$ &  $P = 0.01$ \\
$v = 0$ & $v = 0$ &  $v = 0$ &  $v = 0$ \\
\hline
\end{tabular}
\caption{Initial conditions for two relativistic shocktube problems
  as presented in \cite{marti:03}.
\label{tab:shocktube}
}
\end{table}

\subsection{Sedov Blast Wave}
\label{sec:sedov}

\begin{figure}[t]
\begin{center}
\includegraphics[width=8.0cm]{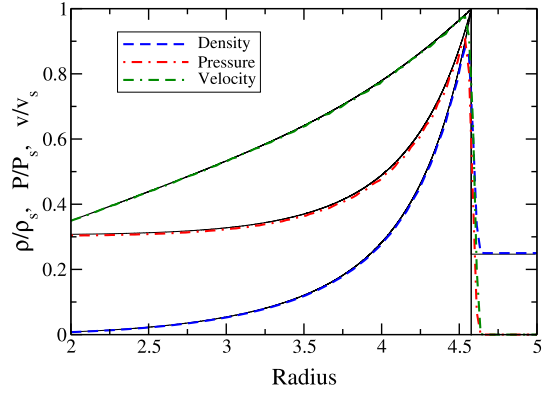} 
\vspace*{-.34cm}
\caption{The Sedov blast wave problem and
  exact solution at $t=0.1$.  Shown are the numerical results with the
  exact solution underlying the various curves of density, pressure
  and velocity.  Both the exact solution and the numerical result are
  normalized to the analytic value at the shock.  ${\rm{\rho_s}} = 4$, ${\rm{P_s}} =
  252.255$ and ${\rm{v_s}} = 13.757$.}\label{fig:sedovcomp}
\end{center}
\end{figure}

The above shocktube tests demonstrated the ability of \code{GR1D} to
capture shocks and solve the special-relativistic hydrodynamic
equations in planar geometry. Here we go back to Newtonian
hydrodynamics and test instead spherical hydrodynamics with Sedov's
blast wave problem \cite{sedov:59}.  For a comparison with a large
number of hydrodynamics codes, we use the initial conditions of
\cite{tasker:08}.  The grid setup is in spherical geometry with
(dimensionless) $r_{max}=10$ and $N=400$ cells which corresponds to
the maximum mesh refinement level used in \cite{tasker:08}. We deposit
a constant specific internal energy into a sphere of radius
$r=0.0875$, corresponding to a total (dimensionless) energy of
$E_o=10^5$, into a background medium of (dimensionless) $\rho_0 = 1$.
We set the background energy density to an insignificant amount and
use a $\Gamma$-law EOS with $\Gamma=5/3$.  Figure~\ref{fig:sedovcomp}
depicts the comparison of our numerical solution with the exact result
for density, velocity and pressure at $t=0.1$ normalized in such a way
that the value of all variables at the shock is 1.  \code{GR1D}
performs very well in the region behind the shock and provides an
adequate, though not perfect, solution near the shock.

In addition to the Newtonian Sedov blast wave problem, we have also
considered its relativistic variant discussed in
\cite{anninos:05}.  These authors used 17 levels
of adaptive mesh refinement (AMR) and we find that the lack of AMR in
\code{GR1D} makes it computationally impossible to adequately resolve
the relativistic Sedov problem. This, however, is not a problem for
the application of \code{GR1D} to the stellar collapse problem, since
the shocks appearing there are only mildly relativistic.

\subsection{Oppenheimer-Snyder Collapse}
\label{sec:oppen}

For the final test problem for which an exact solution exists, we
perform a simulation of the Oppenheimer-Snyder collapse (OSC)
\cite{oppenheimer:39} of a constant-density sphere of pressureless ($P
= 0$) dust. The exact solution of OSC in RGPS spacetime has been laid
out by \cite{petrich:86,gourgoulhon:93b}. We choose $M=M_\odot$,
$R_{\star}=10M_\odot$. We perform the OSC test with the standard
version of \code{GR1D} described in \sref{sec:curvehydro} of this
paper and do not make special adjustments for the code to operate with
$P=0$.  Hence, we set the pressure to a small, but non-zero value,
using a polytropic EOS with $K=10^{-20}$ and $\Gamma=5/3$. In the
artificial atmosphere outside the dust ball, we set the density to
$1\,\mathrm{g\,cm}^{-3}$. We use 9000 equidistant zones to model OSC
with \code{GR1D}.

\begin{figure}[t]
\begin{center}
\hspace*{-0.15cm}\includegraphics[width=7.1cm,clip=]{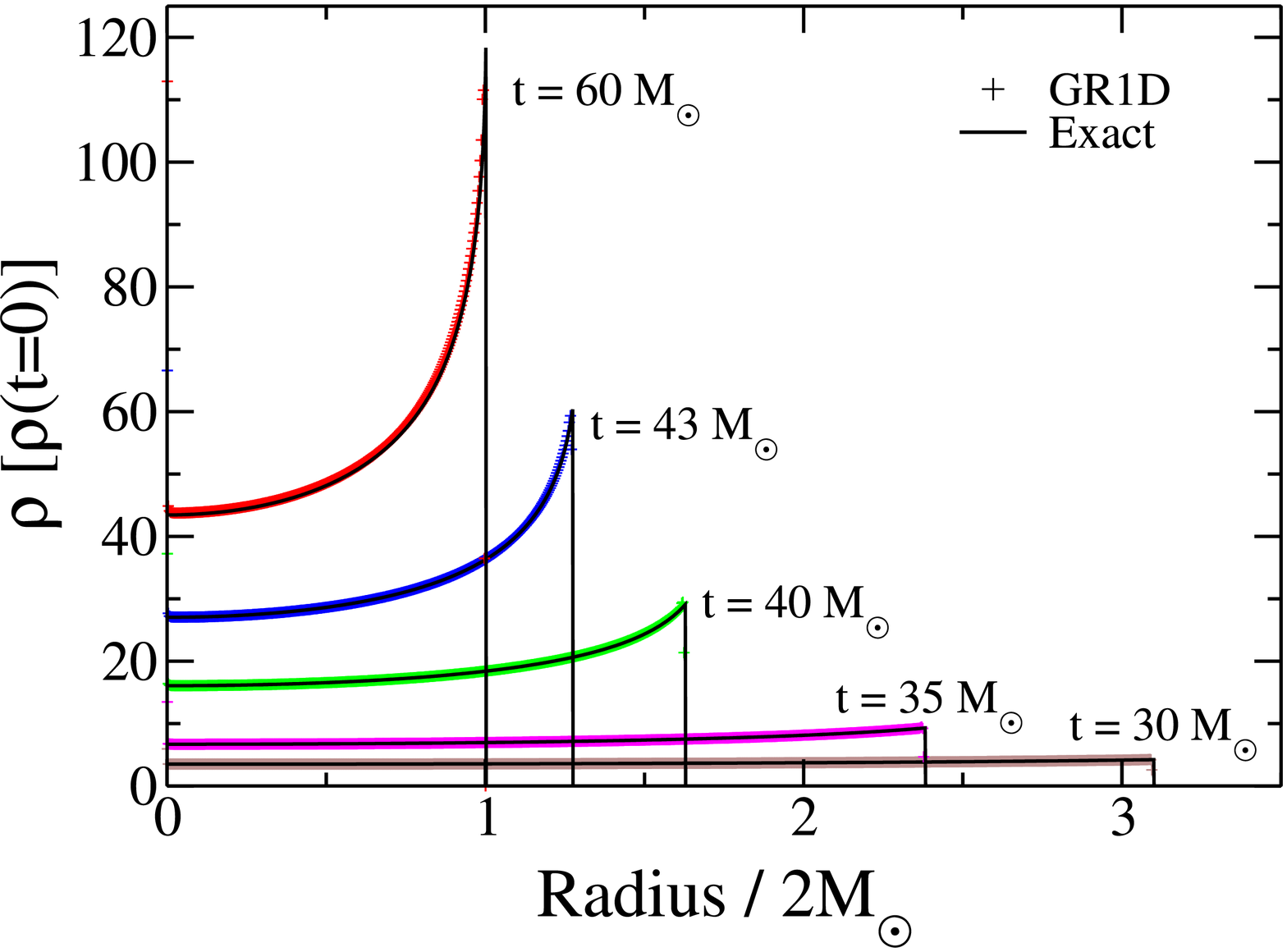} 
\hspace*{-0.7cm}\includegraphics[width=7.1cm,clip=]{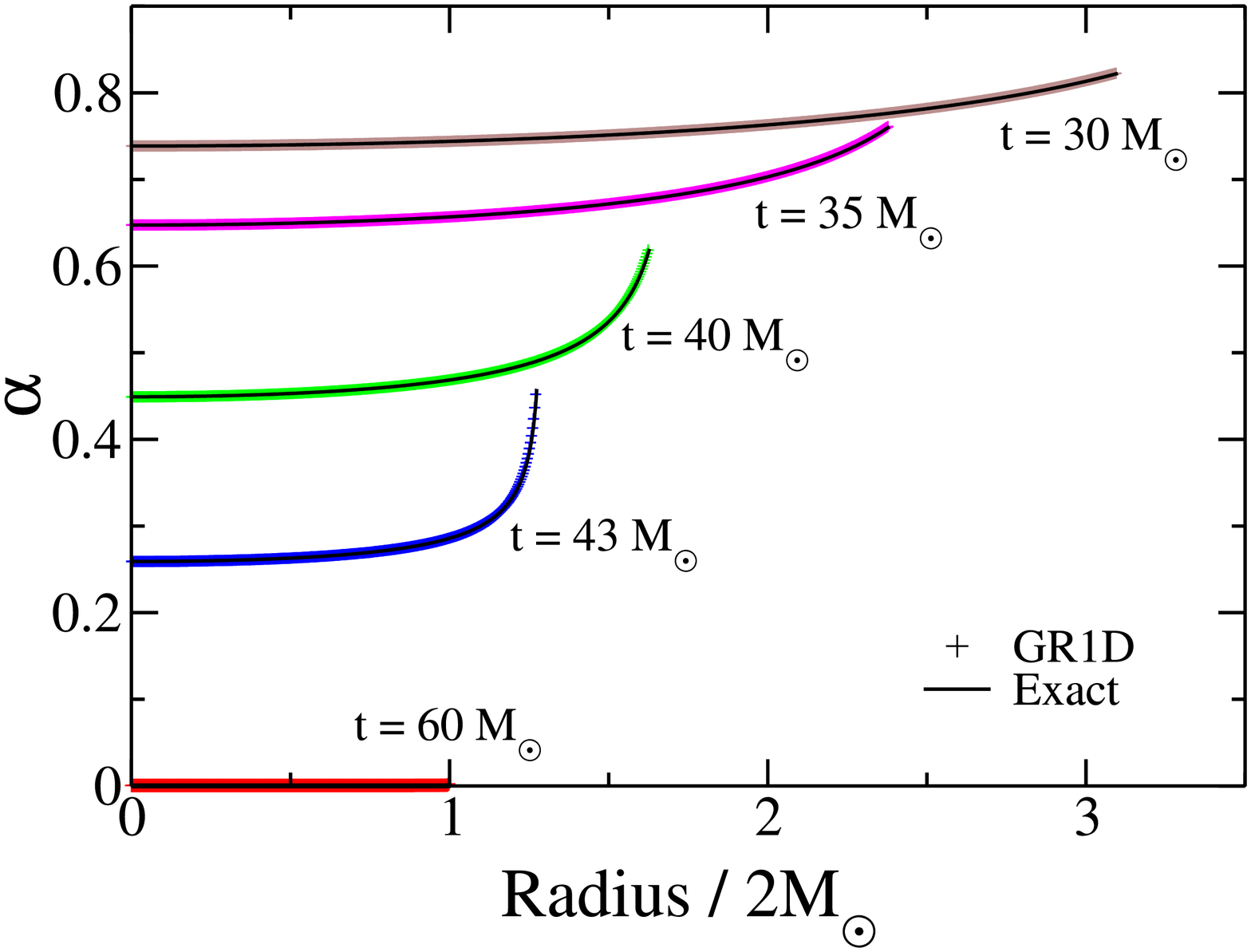}\hspace*{-0.4cm} 
\caption{Oppenheimer-Snyder collapse of a pressureless dust ball.
  Shown are the numerical (plus symbols) and exact (solid lines)
  density (left panel) and lapse (right panel) profiles for various
  times.  The density is normalized to the density at $t=0$.  The
  simulation uses 9000 equally spaced grid points across the domain of
  $20\,M_\odot$.  Initially one solar mass is distributed with constant
  density in a sphere of radius $10\,M_\odot$. For clarity, we show only
  every third data point.}\label{fig:OSCrholapse}
\end{center}
\end{figure}

In \fref{fig:OSCrholapse}, we compare numerical and exact density and
lapse profiles of OSC at $t = 30,\ 35,\ 40,\ 43\ {\rm{and}}\ 60\
M_\odot$.  Following \cite{romero:96}, we normalize the central
density to the value at $t=0$. The overall agreement is
excellent. However, we notice two slight deviations: ${\it{(1)}}$,
near the origin, we observe a small build up of material.  This is
present also in the OSC test of \cite{romero:96} and probably due to
diverging terms near the origin.  We do not notice this effect in our
stellar collapse calculations, most likely because of the stabilizing
effect of the large pressure in the PNS.  ${\it{(2)}}$, at late times
($t>50M_\odot$), the numerical $\alpha$ decreases more slowly then its
exact counterpart and begins to deviate significantly at $\alpha(r=0)
\lesssim 0.001$.  We attribute this to numerical inaccuracies
developing due (\emph{a}) to the metric coefficient $X$ becoming
singular as $R_\mathrm{\star} \to 2 M_\odot$, (\emph{b}) to the extreme
density gradient developing at the surface at late times, and
(\emph{c}) to the fact that we use the standard version of \code{GR1D}
without special adjustments for the OSC problem (as, e.g., made by
\cite{romero:96}).  

\subsection{Hybrid Core Collapse: Convergence}
\label{sec:hybrid}
\begin{figure}[ht]
\begin{center}
\includegraphics[width=9.5cm,clip=]{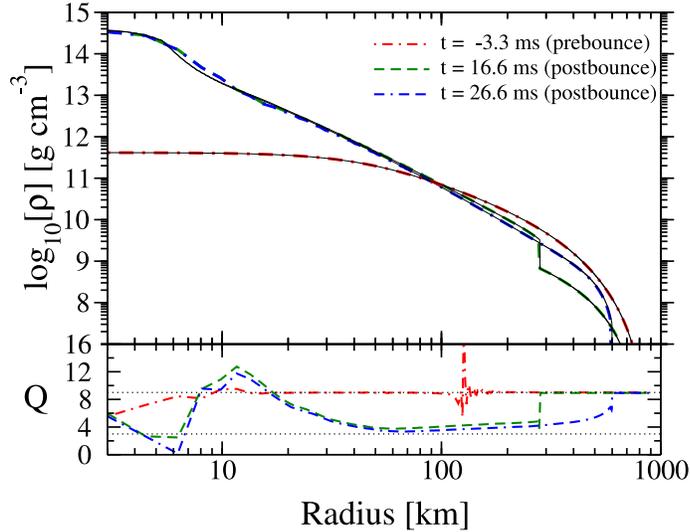} 
\caption{Radial density profiles and self-convergence for core
  collapse using the hybrid EOS. {\textbf{Top:}} Density profiles of the core
  collapse for various times including in the prebounce phase, and
  after the shock has propagated through $\sim 300\ {\rm{and}}\
  600$~km.  We show the low resolution profile (segmented lines) as
  well as the high resolution profile (solid lines) for comparison.
  {\textbf{Bottom:}} Self-convergence of the enclosed gravitational mass,
  $m(r)$.  Dotted lines at Q=3 \& 9 denote expected values for
  1$^\mathrm{st}$ and 2$^\mathrm{nd}$~order convergence.
  }\label{fig:hybridconverge}
\end{center}
\end{figure}

In this section, we present simulations of nonrotating core collapse
and present proof of convergence for \code{GR1D}.  We utilize the
hybrid EOS described in \sref{sec:hybrideos}, taking $\Gamma_1 =
1.28$, $\Gamma_2 = 2.5$, $\Gamma_\mathrm{th} = 1.5$ and
$K=4.935\times10^{14} {\rm{[cgs]}}$. Following \cite{ott:06phd}, we
use as initial data an $n=3$ polytrope with a central density of
$\rho_c = 5 \times 10^{10} {\rm{g}}/{\rm{cm}}^3$ and a $K$ value as
above and initially zero radial velocity.  We simulate the evolution
with \code{GR1D} for equally spaced grids of three different
resolutions ($N_\mathrm{zones} = 500,\ 1500\ {\rm{and}}\ 4500$) to
test the self-convergence of the code.  The self-convergence factor at
convergence order $n$ of a quantity $q$ is given by,
\begin{equation}
Q = {q_1 - q_2 \over q_2 - q_3} = { (dx_1)^n - (dx_2)^n \over (dx_2)^n -
  (dx_3)^n}\,\,,
\label{eq:selfconvergence}
\end{equation}
where $q_i$ is the numerical result from the simulation with the
corresponding resolution and $dx$ is the zone width. For this
convergence test, $dx_1 = 3 dx_2 = 9 dx_3$.  In the lower panel of
\fref{fig:hybridconverge}, we show the self-convergence of
$M_\mathrm{grav}$ at $t=-3.3\,\mathrm{ms}$ (before bounce) as well as at
$t=16.6$~ms and $t=26.6$~ms after bounce.

We generally see the expected $2^\mathrm{nd}$~order convergence (Q=9) in
smooth parts of the flow, but note several interesting features:
{(\emph{1})} before bounce (red, dot-dashed curve) and near $120$~km where
the convergence spikes, the velocity is peaking, causing a reduction
in convergence. {(\emph{2})}, during the postbounce phase, convergence
in the shocked region drops to 1$^\mathrm{st}$~order, this is characteristic
of HRSC schemes in the presence of shocks. {(\emph{3})},
finally, during the postbounce phase for $r<20$~km, the steepness of
the density gradient at the PNS surface and the coarseness of the grid
lead to \emph{local} non-convergence.  We note that the lowest
resolution used here is $d x \sim 2$~km and that
deviations in the density profile compared to higher-resolution
simulations can be seen in the top panel of \fref{fig:hybridconverge}.

\section{Sample Results for a $40$-$M_\odot$ Star}
\label{sec:sampleresults}

In the following simulations we use the single-star, non-rotating,
$M_\mathrm{ZAMS} = 40\ M_\odot$, solar-metallicity presupernova model
of Woosley \& Weaver~\cite{ww:95} (model s40WW95 hereafter).  This
model has an iron core mass of $1.98~M_\odot$. We set up a grid of
$1000$ zones that is logarithmically spaced from $r = 20\,\mathrm{km}$
outward, extending to a radius of $1.15\times10^{5}\,\mathrm{km}$
where the density drops to $200\,\mathrm{g\,cm}^{-3}$.  There is
14.7~$M_\odot$ of baryonic material within this density cutoff. Inside
$r = 20\,\mathrm{km}$, we use an equidistant grid with a spacing of
$100\,\mathrm{m}$.  Such high resolution is necessary to resolve steep
gradients at the PNS surface at late times ($t \gtrsim
0.5\,\mathrm{s}$). Near the origin, we increase the zone size
gradually to $\sim$ 700 m for improved stability but for rotating runs
we find it necessary to maintain the fine grid spacing all the way to
the origin to capture the correct angular velocity profile.

\begin{figure}[t]
\begin{center}
\hspace*{-0.15cm}\includegraphics[width=7.1cm,clip=]{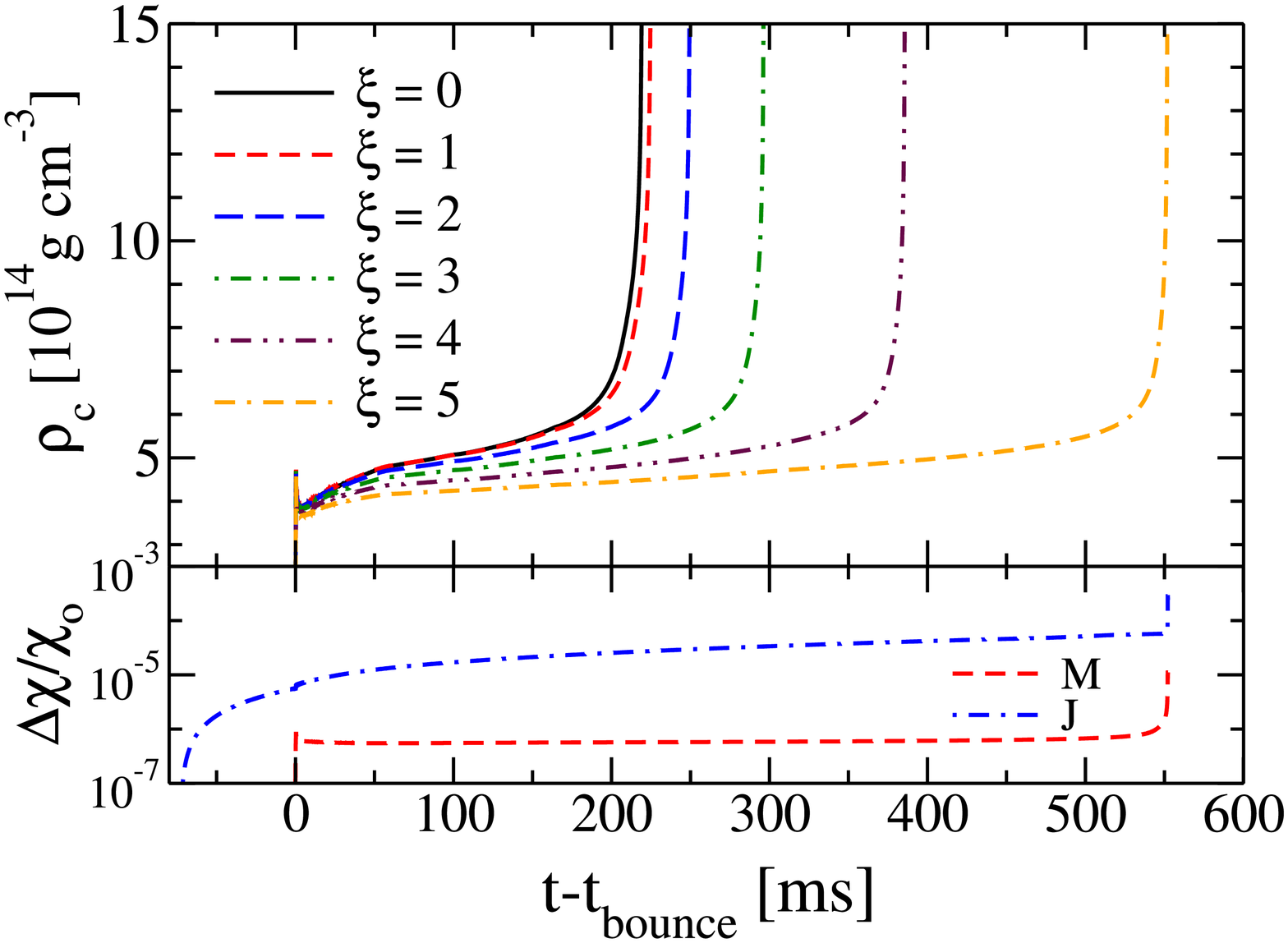} 
\hspace*{-0.4cm}\includegraphics[width=7.1cm,clip=]{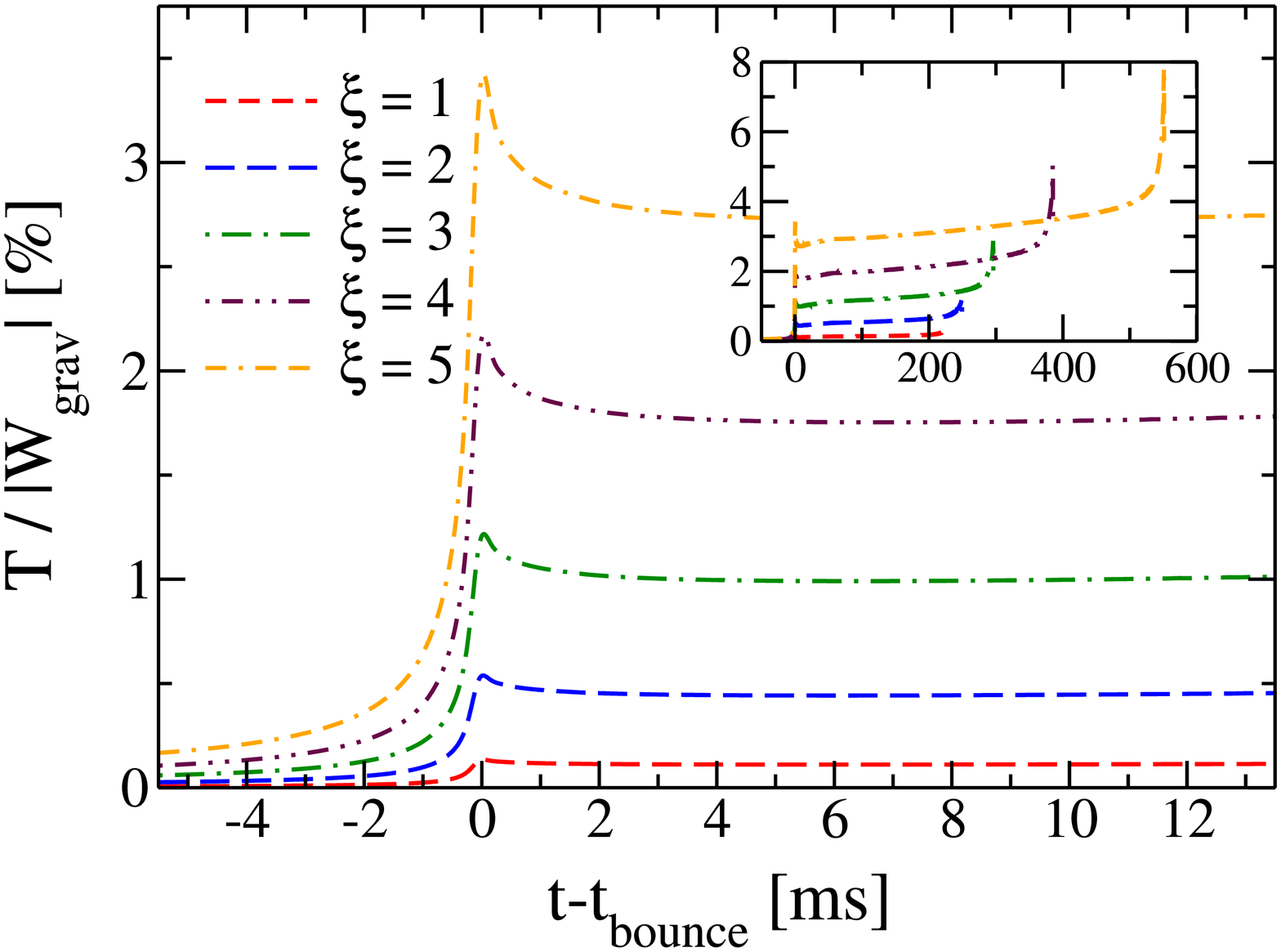}\hspace*{-0.7cm} 
\caption{Black hole formation with rotation and the hybrid
  EOS. {\textbf{Left panel:}} Central densities for various initial
  angular velocities (top) and fractional error of the conserved
  quantities $M_{grav}$ and $J$ (bottom). {\textbf{Right panel:}}
  $T/|W_{\mathrm{grav}}|$ near bounce and (inset) over the entire
  simulation. $\Omega (r)$ is set through
  \eref{eq:omegaofr}.}\label{fig:hybrid_BH} 
\end{center}
\end{figure}

\subsection{Rotating Core Collapse and Black Hole Formation in a $40$-$M_\odot $ Star using the Hybrid EOS}
\label{sec:bhformation}

To show the effects of including rotation and to further demonstrate
the use and usefulness of the hybrid EOS (see \sref{sec:hybrideos})
for exploratory studies, we perform a set of collapse simulations to
black hole formation.  We set $\Gamma_1 = 1.30$, $\Gamma_2 = 2.5$,
$\Gamma_\mathrm{th} = 1.34$ and impose rotation according to the
rotation law (see, e.g., \cite{zwerger:97,ott:06spin})
\begin{equation}
\Omega (r) = \xi {\pi \over 10} \left[1 + \left({r \over
   A}\right)^2\right]^{-1}\,\mathrm{rad\,s}^{-1}\,\,,
\label{eq:omegaofr}
\end{equation} 
where we vary $\xi$ from 0 to 5 and $A$ is a parameter governing the
degree of differential rotation. We choose $A = 1000\,\mathrm{km}$
which leads to roughly uniform rotation within the inner core as
predicted by stellar evolutionary calculations (e.g.,
\cite{heger:00}).  As an additional test of \code{GR1D}, we show in
the lower part of the left panel of \fref{fig:hybrid_BH} the relative
error in total angular momentum and gravitational mass in the most
rapidly spinning simulation. \code{GR1D} conserves angular momentum to
better then one part in $10^{4}$ and $M_\mathrm{grav}$ to one part in $10^{6}$
until the onset of BH formation when the resolution becomes
insufficient to fully resolve the huge gradients in the collapsing
PNS.

We show in the top part of \fref{fig:hybrid_BH} the evolution of the
central density in the simulated models. Due to the choice of
$\Gamma_1$, rotation has little influence on the prebounce dynamics
\cite{dimmelmeier:07}. The hybrid EOS qualitatively captures the
stiffening of the EOS at nuclear density that leads to core bounce.
Owing to the small value of $\Gamma_\mathrm{th}$, the shock stalls
soon after bounce and accretion on the PNS continues. Slowly spinning
models accrete rapidly and collapse to a BH after only 200~ms.
Centrifugal support becomes dynamically relevant in more rapidly
spinning cases, decreasing the accretion rate and delaying BH
formation. The right panel of \fref{fig:hybrid_BH} depicts the
evolution of the rotation parameter $T/|W_{\mathrm{grav}}|$.  Its
systematics are very similar to what has been observed in multi-D
simulations~(e.g., \cite{ott:04,ott:06spin,dimmelmeier:08}).
$T/|W_{\mathrm{grav}}|$ reaches a local maximum at bounce, then
decreases as the PNS reaches its postbounce quasi-equilibrium. New and
not shown before is the evolution of $T/|W_{\mathrm{grav}}|$ near to
BH formation. $T/|W_{\mathrm{grav}}|$ increases only slowly after
bounce (note that, in a calculation with neutrino transport or
leakage, the postbounce $T/|W_{\mathrm{grav}}|$ would increase faster
\cite{ott:06spin}), but near BH formation grows nearly
exponentially during PNS collapse.  Rotation, in particular when it is
strongly differential, can increase the maximum mass of the accreting
PNS (e.g., \cite{baumgarte:00}). We find\footnote{In RGPS, a
  coordinate singularity develops at $R = 2M$ upon BH formation. We
  define here the BH mass to be $M_\mathrm{grav}$ inside the radius
  that corresponds to the maximum $X$. This is an approximation and is
  subject to errors due to our finite resolution grid.}  BH birth
masses of $1.89$-$1.97\,M_\odot$ for the set of rotating hybrid-EOS
models considered here.  This increase in the maximum mass is modest,
primarily because our PNS cores are rather uniformly spinning (in
agreement with \cite{ott:06spin,dimmelmeier:08}).  We point out that
our present treatment does not consider angular momentum
redistribution by multi-dimensional effects or effective viscosity
which may be present in realistic systems (see, e.g.,
\cite{ott:07prl,thompson:05} and references therein).

Finally, we note that for the nonrotating ($\xi = 0$), the evolution
with \code{GR1D} continues until a central value of the lapse function
of $3\times10^{-10}$ and a maximum value of $\sqrt{g_{rr}} = X$ of
$\sim 21.1$. These are excellent values in comparison to previous
studies on BH formation in RGPS~\cite{gourgoulhon:91,gourgoulhon:93b}.
In the rotating case, the evolution terminates somewhat earlier due
primarily to numerical issues near the origin at very large $v_\varphi$.

\subsection{Nonrotating Collapse and Black Hole Formation with
Neutrino Leakage/Heating in a $40$-$M_\odot $ Star}
\label{sec:s40_leakage}

\begin{figure}[t]
\begin{center}
\hspace*{-0.15cm}\includegraphics[width=7.1cm,clip=]{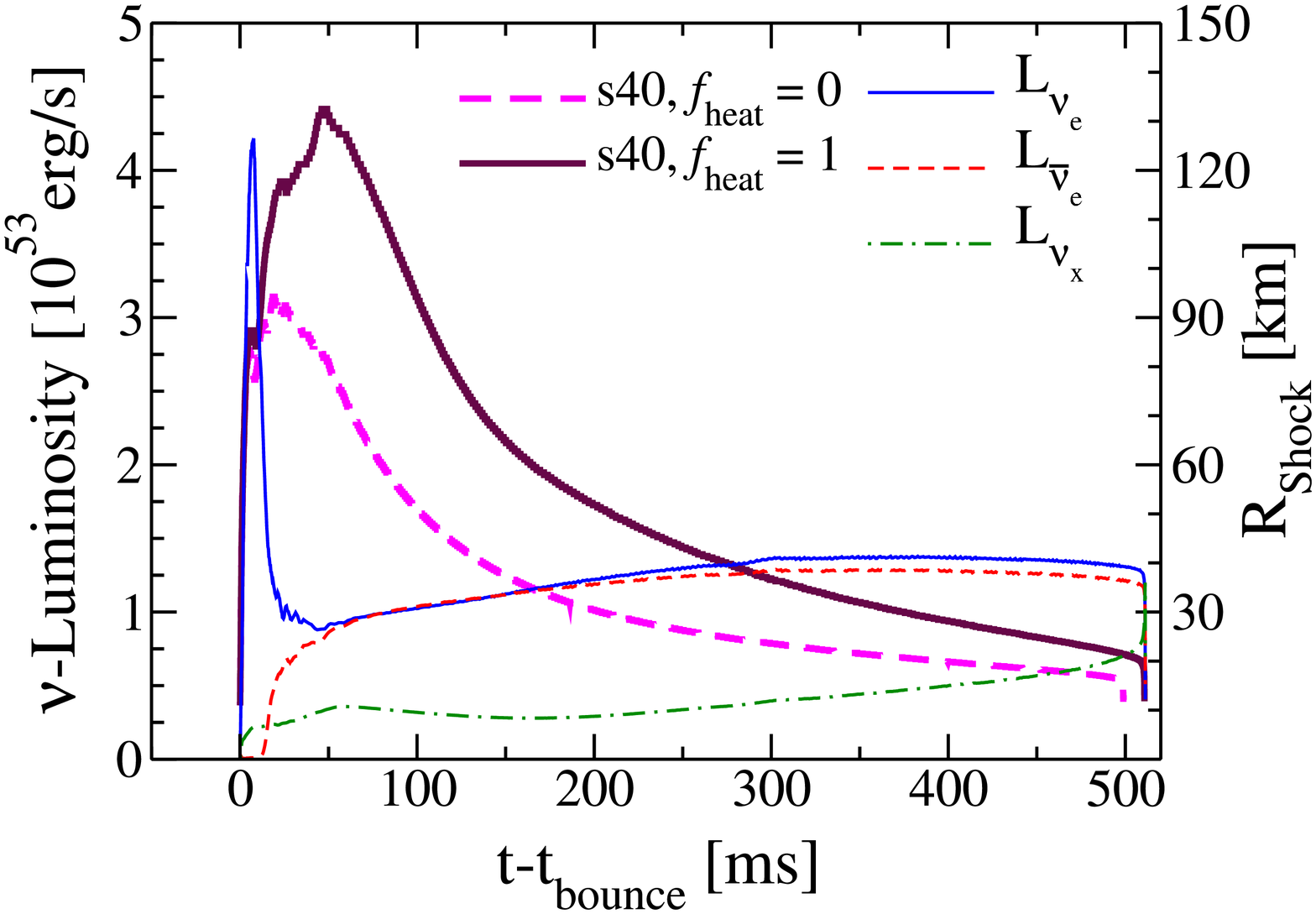} 
\hspace*{-0.3cm}\includegraphics[width=7.1cm,clip=]{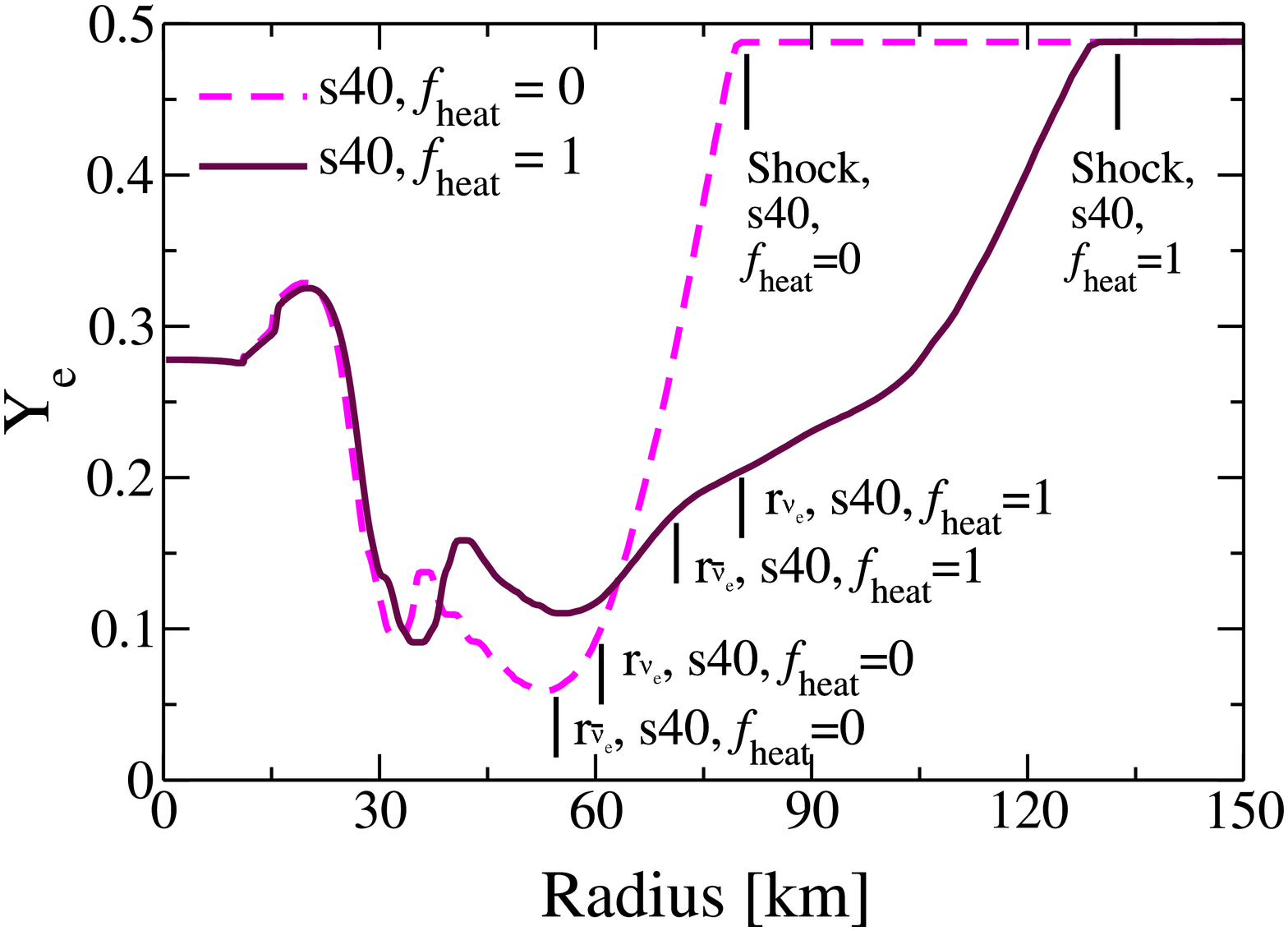}\hspace*{-0.8cm} 
\caption{{\bf Left panel}: Shock radius (thick line, right ordinate) and
  neutrino luminosities (thin lines, left ordinate) as a function of
  postbounce time in a nonrotating leakage+heating
  ($f_\mathrm{heat}=1$) simulation with the 40$M_\odot$-model of
  \cite{ww:95} run with the LS180 EOS.  Shown also is the shock radius
  evolution (dashed thick lines) in a simulation without heating
  $f_\mathrm{heat}=0$.  {\bf Right panel:} $Y_e$ profiles of both
  simulations at $50\,\mathrm{ms}$ after bounce, corresponding to the
  maximum shock radius of the $f_{heat}=1$ simulation. Shock radii,
  electron and anti-electron neutrino neutrinospheres are marked for
  both the $f_{heat}=1$ and $f_{heat}=0$
  simulations.}\label{fig:s40shockandye}
\end{center}
\end{figure}

In this section we show example results employing \code{GR1D}'s
leakage/heating scheme and finite-temperature EOS.  We use the s40WW95
progenitor and the LS180 EOS\footnote{The lower bound on our EOS
  tables is $1000\,\mathrm{g\, cm}^{-3}$, we bring the outer boundary
  into $\rho = 2000\,\mathrm{g\, cm}^{-3}$ for this example.},
$Y_e(\rho)$ parameterization pre-bounce, our standard leakage/heating
scheme after bounce, and no rotation. We show results for both
$f_\mathrm{heat} = 0$ (losses only) and $f_\mathrm{heat} = 1$.  In
\fref{fig:s40shockandye}, we compare the shock radii of these two runs
and neutrino luminosities of the $f_{\mathrm{heat}}=1$ run (left
panel) as well as the $Y_e$ radial profiles at $50$~ms after bounce
(right panel).  We note that the total luminosity is $L_{\nu_e} +
L_{\bar{\nu}_e} + 4L_{\nu_\mu}$ and is corrected for redshift through
\eref{eq:nulumCF_heat} with $r=\infty$, but, nevertheless, is somewhat
higher (up to $\sim$~20\%) than predicted by full Boltzmann
radiation-hydrodynamics calculations using the same
progenitor~\cite{fischer:09a,sumiyoshi:07}.  The time until BH
formation in the case of $f_\mathrm{heat}=1$ is $t_\mathrm{BH} =
511$~ms and the baryonic mass inside the shock of the last stable
configuration is 2.25~$M_\odot$.  We compare this to two other studies
of BH formation in 1D with the same progenitor model and EOS, but with
two different implementations of GR Boltzmann neutrino
transport. These studies are Fischer~et~al.~\cite{fischer:09a} who
found $t_\mathrm{BH}=435.5$~ms and 2.196~$M_\odot$ and
Sumiyoshi~et~al.~\cite{sumiyoshi:07}, who found $t_\mathrm{BH}=560$~ms
and 2.1~$M_\odot$.  Our result is very close to these more accurate
studies which gives us confidence in the robustness of the
heating/leakage scheme in \code{GR1D}.

The right panel of \fref{fig:s40shockandye} depicts the $Y_e$ profiles
at $50\,\mathrm{ms}$ after bounce. The characteristic trough in $Y_e$
behind the shock is captured by our leakage/heating scheme, but we
find that our simple heating scheme converts too many of the postshock
neutrons back to protons at early times, leading to too high values of
$Y_e$ in the lower postshock region between $\sim 30-60\,\mathrm{km}$.

To conclude this section,  we note that, due to the computational efficiency of
our scheme, each of our simulations took only $\sim$~6 CPU hours
from iron core collapse through BH formation on one core of
an Intel Xeon X5550 (Nehalem) machine.

\section{Summary and Concluding Remarks}
\label{sec:summary}

In this paper, we have presented the details of our new open-source
Eulerian 1.5D GR hydrodynamics code \code{GR1D}.
\code{GR1D} is intended primarily for the simulation of stellar
collapse to neutron stars and black holes and, for the first time in
the 1D GR context, includes an approximate way of accounting for
stellar rotation consistent with that used in state-of-the-art
calculations of stellar evolution (e.g., \cite{heger:00}).  Using this
scheme, we have presented rotating long-term postbounce
simulations towards black hole formation using a $40$-$M_\odot$
supernova progenitor model and showed how the simple analytic hybrid
EOS can be used to capture many qualitative aspects of this phenomenon.

As we have demonstrated in this paper, \code{GR1D} performs well in
standard tests and, despite its simplified neutrino leakage/heating
scheme, still yields overall results in the case of failing
core-collapse supernovae and black hole formation that measure up 
qualitatively and to some extent also quantitatively to
those obtained with full Boltzmann neutrino transport in 1D
Lagrangian codes \cite{fischer:09a,sumiyoshi:07}.

Many 1D GR (radiation)-hydrodynamics formulations have been presented
in the past $\sim$$50$~years. Yet, there is presently no open-source
1D GR stellar collapse code available to the broader community.  The
primary motivation driving the development of \code{GR1D} is the need
for such an open-source code that may be used as a codebase,
benchmark, and testbed for improved modeling technology to be included
in multi-D GR codes addressing core-collapse supernova explosions, but
also failing core-collapse supernovae, black hole formation, and the
post-merger evolution of binary neutron-star and neutron-star -- black
hole coalescence.  Equipped with an approximate neutrino-leakage
scheme to capture the key effects associated with neutrino heating and
cooling, the version of \code{GR1D} discussed in
this paper is a solid starting point for the next generation of 
astrophysically-relevant multi-D GR simulations.

The current limitations of \code{GR1D} due to its gray leakage and
simplified heating scheme are obvious.  We will continue to
develop and improve \code{GR1D} and intend to include as a next step
energy-dependent radiation transport in the multi-group
flux-limited diffusion approximation (MGFLD) and/or in the isotropic
diffusion source approximation (IDSA, \cite{liebendoerfer:08}).

\section*{Acknowledgements}
We thank the Niels Bohr International Academy for hosting the
Microphysics in Computational Relativistics Astrophysics (MICRA)
workshop in August 2009 at which much of the work presented here was
inspired. It is a pleasure to thank J.-M. Ib\'an\~ez for helpful
advice, for providing the original version of the code of
Romero~et~al., and for furnishing a copy of Romero's dissertation.  We
are indebted to M.~Duez for very valuable help with the derivation of
the neutrino source terms.  We are furthermore happy to acknowledge
helpful exchanges with W.~D.~Arnett, A.~Burrows, P. Cerd\'a-Dur\'an,
H.~Dimmelmeier, T.~Fischer, E.~Gourgoulhon, I.~Hawke, J.~Lattimer,
L.~Lehner, M.~Liebend\"orfer, E.~Livne, C.~Meakin, S.~Noble,
A.~Perego, C.~Pethick, E.~S. Phinney, E.~Schnetter, S.~Scheidgger,
Y.~Sekiguchi, and S.~Teukolsky.  This work is supported by the
National Science Foundation under grant numbers~AST-0855535 and
OCI-0905046.  EOC is supported in part through a post-graduate
fellowship from the Natural Sciences and Engineering Research Council
of Canada (NSERC) and NASA ATP grant NNX07AH06G.  We wish to thank
Chris Mach for support of our group servers at TAPIR on which much of
the code development and testing was carried out.  Results presented
in this article were obtained through computations on the NSF Teragrid
under grant TG-MCA02N014, on machines of the Louisiana Optical Network
Initiative under grant LONI\_NUMREL04, and at the National Energy
Research Scientific Computing Center (NERSC), which is supported by
the Office of Science of the US Department of Energy under contract
DE-AC03-76SF00098.

{\scriptsize \setlength{\parskip}{0.1cm} }

\appendix

\section{Evolution Equation Derivation}
\label{sec:appendix}

In this appendix we derive the evolution equations for the conserved
variables $D, DY_e, S^r, S_\phi$ and $\tau$ used in \code{GR1D} and
presented in \sref{sec:grhydro} and \ref{sec:rotation}.  \code{GR1D}
uses the spherically symmetric metric $g_{\mu \nu} = \mathrm{diag}
(-\alpha^2, X^2, r^2, r^2\sin^2{\theta})$ with $\alpha =
\exp{(\Phi(r,t))}$ with $\Phi(r,t)$ defined through \eref{eq:phi}, $X
= (1 - {2m(r,t) \over r})^{-1/2}$ where $m(r,t)$ is the enclosed
gravitational mass at coordinate radius $r$.  We assume the matter to
be a perfect fluid described by a mass current density of $J^\mu = \rho
u^\mu$ and a stress-energy tensor, $T^{\mu \nu} = \rho h u^\mu u^\nu +
g^{\mu \nu}P$ where $\rho$ is the rest mass density, $P$ is the fluid
pressure, $h = 1 + \epsilon + P/\rho$ is the specific enthalpy with
$\epsilon$ the specific internal energy; $u^\mu = (W/\alpha,W v
/X,0,0)$ is the fluid 4-velocity (without taking into account
rotation) with $W=1 / \sqrt{1 - v^2}$ is the Lorentz factor and $v$ is
the physical radial velocity. 

While evaluating the covariant derivative of the stress-energy tensor
and matter current density, we make use of the following formula,
\begin{equation}
\nabla_\mu J^\mu = {1 \over \sqrt{-g}} \left(\sqrt{-g}
    J^\mu\right)_{,\mu}
\end{equation}
and
\begin{equation}
 \nabla_\mu T^{\mu \nu} = {1 \over \sqrt{-g}} \left(\sqrt{-g}
    T^{\mu \nu}\right)_{,\mu} + \Gamma^\nu_{\alpha \mu}T^{\mu \alpha}\,,
\end{equation}
where $\sqrt{-g} = \alpha X r^2$ is the determinant of the metric and
${\Gamma^\nu}_{\alpha \mu}$ are Christoffel symbols and are defined
through derivatives of the metric,
\begin{equation}
{ {\Gamma}^{\nu} }_{\alpha \mu} = \frac {1}{2} g^{\nu \beta}
( g_{\mu \beta, \alpha} + g_{\alpha \beta, \mu} - g_{\alpha \mu, \beta})\,.
\end{equation}

For our metric, all non-zero Christoffels are given in
Table~\ref{tab:christoffels}, ${\Gamma^\nu}_{\alpha \mu}$ is symmetric
in the last two indices, duplicates are omitted.

\begin{table}[ht]
\centering
\begin{tabular}{|rl|rl|}
\hline
${\Gamma^{t}}_{tt}$ =  &$\partial_t \phi(r,t)$ & ${\Gamma^{r}}_{\theta \theta}$ =  &$-{r \over X^2}$\\[3pt]
${\Gamma^{t}}_{tr}$ =  &$\partial_r \phi(r,t)$ & ${\Gamma^{r}}_{\phi \phi}$ =  &$- {r \sin^2{\theta} \over X^2}$ \\[3pt]
${\Gamma^{t}}_{rr}$ =  &$\alpha^{-2}{X^4 \over r}\partial_t m(r,t)$  & ${\Gamma^{\theta}}_{r \theta}$ =  &${1 \over r}$\\[3pt]
${\Gamma^{r}}_{tt}$ =  &${\alpha^2 \over X^2}\partial_r \phi(r,t)$ & ${\Gamma^{\theta}}_{\phi \phi}$ =  &$-\sin{\theta}\cos{\theta}$\\[3pt]
${\Gamma^{r}}_{tr}$ =  &${X^2 \over r}\partial_t m(r,t)$ & ${\Gamma^{\phi}}_{r \phi}$ =  &${1 \over r}$\\[3pt]
${\Gamma^{r}}_{rr}$ =  &${X^2 \over r}(\partial_r m(r,t) - {m(r,t) \over r})$ & ${\Gamma^{\phi}}_{\theta \phi}$ =  &${\cos{\theta} \over \sin{\theta}}$ \\[3pt]
\hline
\end{tabular}
\caption{Connection coefficients.} \label{tab:christoffels}
\end{table}

It is useful to note the following derivatives needed in the
derivation of the evolution equations:
\begin{eqnarray}
\label{eq:derivatives1}
\partial_r \Phi(r,t)& =& X^2\left[{m
    \over r^2} + 4\pi r (P + \rho h W^2 v^2)\right] \,,\\
\label{eq:derivatives2}
 \partial_r X &=&X^3 \left[{\partial_r m \over r} - {m \over
     r^2}\right]\,, \\
\label{eq:derivatives3}
\partial_tX &=& X^3 {\partial_t m \over r}\,,\\
\label{eq:derivatives4}
 \partial_r m& =& 4 \pi r^2(\rho h W^2 - P)\,, \\
\label{eq:derivatives5}
 \partial_t m &=& - 4 \pi r^2 {\alpha \rho h W^2v \over X} \,.
\end{eqnarray}

\subsection{Source Terms}
\label{sec:appsource}

The evolution equations follow from $\nabla_\mu J^\mu = 0$ and
$\nabla_\mu T^{\mu \nu} = 0$. Since we treat neutrinos through a
leakage scheme, we add in neutrino source terms explicitly to the RHS
of these equations.  The neutrino physics of \code{GR1D} occurs in the
rest frame of the fluid; in this frame the energy and lepton rates are
calculated with the neutrino leakage scheme, $Q^0_E$ and $R^0_{Y_e}$
are given in \eref{eq:leakagerates}.  Momentum exchange in the fluid
rest frame is taken into account approximately via $Q_M^0 =
-\frac{\partial P_\nu}{\partial r}$ where the gradient is evaluated
numerically in the coordinate frame. This introduces a slight
inconsistency, since in a full radiation-transport treatment the
momentum transfer is computed fully locally via the second angular
moment of the local neutrino radiation intensity \cite{mueller:10}.

By writing the evolution equations in the comoving orthonormal frame
of the fluid (fluid rest frame, [FRF]) with 4-velocity $\vec{u} =
(1,0,0,0)_{\rm{FRF}}$ and unit radial normal $\vec{n} =
(0,1,0,0)_{\rm{FRF}}$ and expressing them as frame-independent tensor
equations we can derive expressions for the evolution equations in any
frame.  For the lepton fraction,

\begin{eqnarray}
\nonumber
\partial_t (\rho Y_e) &=& R^0_{Y_e}\, ,\\
\nonumber
\partial_t (\rho Y_e u^t ) & =& R^0_{Y_e}\, , \\
\nonumber
\partial_\mu (\rho Y_e u^\mu ) & =& R^0_{Y_e}\, , \\
\nabla_\mu (\rho Y_e u^\mu ) & = & R^0_{Y_e}\, .\label{eq:FRFYeevolution}
\end{eqnarray}

We write the energy and momentum source terms in the fluid rest frame
as a 4-vector, $\vec{q} = (Q^0_E, Q^0_M, 0, 0)_{\rm{FRF}}$ or in
frame-independent notation, $Q^0_E \vec{u} + Q^0_M\vec{n}$.  In the
fluid rest frame, the evolution equations for energy and momentum
become,
\begin{equation}
\partial_t T^{tt} = Q^0_E = q^t ,
\end{equation}
and
\begin{equation}
\partial_t T^{tr} = Q^0_M = q^r ,
\end{equation}
or in frame-independent tensor notation,
\begin{equation}
\nabla_\mu T^{\mu \nu}  = q^\nu\,. \label{eq:FRFEMevolution}
\end{equation}

For the evolution equations, we must transform $\vec{q}$ from the
fluid rest frame, to the coordinate frame (CF) of \code{GR1D}.  In a
general frame $\vec{n}$ is a vector that is both {\emph{i)}}
normalized and {\emph{ii)}} orthogonal to $\vec{u}$.  In the CF of
\code{GR1D}, where $\vec{u}$ is the 4-velocity, these two conditions
(along with the assumption of spherical symmetry) on $\vec{n}$ give
$\vec{n} = \left(W v / \alpha,W / X,0,0\right)_{\rm{CF}}$.
$\vec{q}$ in the CF then becomes $\vec{q} = \left({W\over
\alpha}(Q^0_E + vQ^0_M),{W\over X} (vQ_E^0 +
Q^0_M),0,0\right)_{\rm{CF}}$.  This can also be derived via a
Lorentz transformation.  In principle, non-zero rotation will
give rise to source terms for the $\phi$-momentum evolution through
$q^\phi$ and modify the radial source terms $q^r$.  In consideration
of the significant approximations already present in both our neutrino
leakage scheme and in our treatment of rotation, we neglect the
influence of rotation on the source terms. This is justified as long
as $v_\varphi \ll c$.

\subsection{\code{GR1D} Evolution Equations}
\label{sec:appevolution}

In the coordinate frame of \code{GR1D} where $u^\mu = (W/\alpha,
Wv/X,0,0)$, the continuity equation, $\nabla_\mu J^\mu=0$ gives the
evolution of the rest mass density,
\begin{eqnarray}
\nonumber
\nabla_\mu(\rho u^\mu) & = & 0\, , \\
\nonumber
{1 \over \sqrt{-g}} \left[\partial_t \left(\sqrt{-g}{\rho W \over \alpha}\right)
 + \partial_r\left(\sqrt{-g} {\rho W v \over X}\right)\right] &=& 0\, , \\
\partial_t(D) + {1 \over r^2}\partial_r\left({\alpha r^2 \over X} D v\right)
 &=&0\, .\label{eq:restmassevolution}
\end{eqnarray}

The evolution of the electron fraction $Y_e$ follows a similar
derivation but contains a source term from the neutrino leakage
scheme.  In the coordinate frame of \code{GR1D}
\eref{eq:FRFYeevolution} becomes,
\begin{eqnarray}
\nonumber
\nabla_\mu(\rho Y_e u^\mu) & = & R^0_{Y_e}\, , \\[3pt]
\nonumber
{1 \over \sqrt{-g}} \left[\partial_t \left(\sqrt{-g} {\rho W Y_e \over \alpha}\right)
 + \partial_r\left(\sqrt{-g} {\rho W Y_e v \over X}\right)\right] &=& R^0_{Y_e}\, , \\[3pt]
\nonumber
{1 \over \alpha X} \left[ \partial_t \left(X\rho W Y_e\right) + {1 \over
    r^2}\partial_r\left({\alpha r^2 \over X} X\rho W Y_e v\right)\right] &=&
R^0_{Y_e}\, ,\\[3pt]
\partial_t(DY_e) + {1 \over r^2}\partial_r\left({\alpha r^2 \over X} DY_ev\right)
 &=&\alpha X R^0_{Y_e}\, .
\end{eqnarray}

The momentum evolution equation for \code{GR1D}
is obtained by evaluating \eref{eq:FRFEMevolution} with $\nu = r$.
\begin{eqnarray}
  \nonumber
  \fl \nabla_\mu T^{\mu r} &=& q^r\, ,\\[3pt]
  \nonumber
  \fl \left(\sqrt{-g}\ T^{\mu r}\right)_{, \mu} &=&\sqrt{-g}\ q^r
  -\sqrt{-g}\ \Gamma^r_{\nu \mu}T^{\mu \nu} \, ,\\[3pt]
  \nonumber
  \fl\partial_t\left( \rho h W^2 v \right)
  + {1 \over r^2}\partial_r\left({\alpha r^2 \over X} \left(\rho h W^2
      v^2 + P\right)\right) &=& \alpha X q^r - \alpha X\big(
  \Gamma^r_{\nu t} T^{t \nu} + \Gamma^r_{\nu r}T^{r \nu} \\
  \nonumber
  &&\hspace*{-1cm} + \Gamma^r_{\nu \phi}T^{\phi \nu} + \Gamma^r_{\nu \theta}T^{\theta \nu}\big)\,,\\[3pt]
  \nonumber
  \fl\partial_t\left(S^r \right)
  + {1 \over r^2}\partial_r\left({\alpha r^2 \over X}
    \left(S^rv+P\right)\right) &=& \alpha X q^r - \alpha X \big(
  \Gamma^r_{t t} T^{t t} + \Gamma^r_{r t}T^{t r }\\
  \nonumber
  &&\hspace*{-1cm} + \Gamma^r_{t r}T^{r t} + \Gamma^r_{r r }T^{r r}+
  \Gamma^r_{\phi \phi}T^{\phi \phi} + \Gamma^r_{\theta
    \theta}T^{\theta \theta}\big)\,,\\[3pt]
  \nonumber
  \fl\partial_t\left(S^r \right)
  + {1 \over r^2}\partial_r\left({\alpha r^2 \over X}
    \left(S^rv+P\right)\right) &=& - \alpha X \Bigg[ 2 {X
    \over r } {\rho h W^2 v \over \alpha X}\partial_t m \\
  \nonumber
  &&\hspace*{-1cm}+ {X^2 \over r}\left(\partial_r m - {m \over
      r}\right)\left({\rho h W^2v^2 + P \over X^2}\right) \\
  \nonumber
  &&\hspace*{-1cm}  - {2 P \over X^2 r}  + {\alpha^2
    \over X^2}{\rho h W^2 - P \over \alpha^2} \partial_r \Phi \Bigg]+ \alpha X q^r \,, \\
  \nonumber
  \fl \partial_t\left(S^r\right) +{1 \over r^2} \partial_r\left[ {\alpha r^2 \over X}(S^rv +
    P)\right]  &=& \alpha X\bigg[(S^rv-\tau -D) \left(8\pi r P + {m \over
      r^2}\right) \\
  \fl && \hspace*{-1cm}+ {P m \over r^2} + {2 P \over X^2 r}\bigg] +\alpha W (vQ^0_E+Q^0_M)\,. \label{eq:momentumevolution}
\end{eqnarray}
where in the last step we have reorganized the source terms to the
form of \cite{romero:96} using the derivatives defined in
\eref{eq:derivatives1}-\eref{eq:derivatives5}.  If non-zero, $u^\phi =
Wv_\varphi / r$ leads to an additional term ($\alpha \rho h W^2
v_\varphi^2 \sin(\theta)^2 / Xr$) arising through $\Gamma^r_{\phi
  \phi} T^{\phi \phi}$ on the RHS of \ref{eq:momentumevolution},
averaging this term over the spherical shell gives $2/3\ \alpha \rho h
W^2 v_\varphi^2 / Xr$.  When rotation is included, the evolution
equation for $S_\phi=\rho h W^2 v_\varphi r$ is,
\begin{eqnarray}
\nonumber
\fl \nabla_\mu T^{\mu}_\phi &=&0\, ,\\[3pt]
\nonumber
\fl \left(\sqrt{-g}\ T^\mu_\phi\right)_{,\mu}
&=&\sqrt{-g}\ \Gamma^\nu_{\phi \mu}T^\mu_\nu\,, \\[3pt]
\nonumber
\fl \partial_t\left(\alpha X r^2 g_{\phi \phi}T^{t
    \phi}\right) + \partial_r\left(\alpha X r^2 g_{\phi
    \phi}T^{r \phi}\right) &=& \sqrt{-g} \left(\Gamma^r_{\phi \phi}
  T^\phi_r + \Gamma^\phi _{\phi r}T^r_\phi\right)\,,\\[3pt]
\nonumber
\fl \partial_t(X \rho h W^2 v_\varphi r) + {1
    \over r^2}\partial_r\left({\alpha r^2 \over X} \rho h W^2
    v_\varphi r  v  X\right)  &=& 0\,,\\[3pt]
\nonumber
\partial_t(S_\phi) + {1
    \over r^2}\partial_r\left({\alpha r^2 \over X} S_\phi v\right) &=&
  {\rho h W^2 v_\varphi r \over X}\left(-\partial_tX - {\alpha v \over
      X}\partial_r X\right)\,, \\ [3pt]
\partial_t(S_\phi) + {1
  \over r^2}\partial_r\left({\alpha r^2 \over X} S_\phi v\right) &=& \alpha \rho h W^2 v_\varphi v X \left(4 \pi r^2
  P + {m \over r}\right)\,.
\end{eqnarray}

The energy evolution equation for \code{GR1D} is derived by taking
$\nu=t$ in \eref{eq:FRFEMevolution},
\begin{eqnarray}
\nonumber
  \fl \nabla_\mu T^{\mu t} &=& q^t\, ,\\[3pt]
\nonumber
\fl \left(\sqrt{-g}\ T^{\mu t}\right)_{,\mu}
&=&\sqrt{-g}\ q^t-\sqrt{-g}\ \Gamma^t_{\nu \mu}T^{\mu \nu}\,, \\[3pt]
\nonumber
\fl \partial_t\left({X \over \alpha} (\rho h W^2 - P)\right)
+ {1 \over r^2}\partial_r\left({\alpha r^2 \over X} \rho h W^2 v {X
    \over \alpha}\right)
&=& \alpha X q^t - \alpha X \left(\Gamma^t_{t \mu}T^{\mu t} +
  \Gamma^t_{r \mu}T^{\mu r}\right)\,,\\[3pt]
\nonumber
\fl {X \over \alpha}\left[\partial_t\left( \tau + D\right)
+ {1 \over r^2}\partial_r\left({\alpha r^2 \over X} S^r \right)\right]
&=& \alpha X q^t - \alpha X \big(\Gamma^t_{t t}T^{t t} +2\Gamma^t_{t
  r}T^{r t} \\
\nonumber
\fl &&\hspace*{-1cm} +\Gamma^t_{r r}T^{r r}\big) - (\rho h W^2 -
P)\partial_t\left({X \over \alpha}\right) \\
\nonumber
\fl && \hspace*{-1cm}- {\alpha \rho h W^2 v
  \over X} \partial_r\left({X
  \over \alpha}\right)\,,\\[3pt]
\nonumber
\fl \partial_t\left( \tau + D\right)
+ {1 \over r^2}\partial_r\left({\alpha r^2 \over X} S^r \right)
&=& \alpha^2 q^t\,, \\[3pt]
\fl \partial_t\left( \tau \right)
+ {1 \over r^2}\partial_r\left({\alpha r^2 \over X} (S^r-Dv) \right)
&=& \alpha W (Q^0_E + vQ^0_M) \,.
\end{eqnarray}
where in the last step we use the continuity equation
\eref{eq:restmassevolution} to subtract out the evolution of the rest
mass density, obtaining the evolution equation for $\tau$.  A non-zero
$u^\phi$ does not contribute source terms to this evolution equation.

\section{Neutrino Luminosities}
\label{sec:appendixlum}

The luminosity computed from the neutrino leakage scheme is derived in
the rest frame of the fluid.  We require knowledge of the neutrino
luminosity as measured by an observer at rest in the coordinate frame
to determine {\emph{i)}} the luminosity measured by an observer at
rest at infinity and {\emph{ii)}} the luminosity in the fluid rest
frame at some other coordinate radius for our neutrino heating scheme.
We derive these relationships by assuming the neutrinos are emitted
radially in the fluid rest frame with energy $E^{\rm{FRF}}$.

In the fluid rest frame (FRF), the 4-momentum of the (massless)
neutrino is $p^a = (E^{\rm{FRF}},E^{\rm{FRF}},0,0)_{\rm{FRF}}$.  We
use the orthonormal tetrad in \ref{sec:appsource}, in the fluid frame,
$\vec{u}= \vec{e}_0= (1,0,0,0)_{\rm{FRF}}$ and $\vec{n} = \vec{e}_1 =
(0,1,0,0)_{\rm{FRF}}$, in the coordinate frame (CF), $u^\beta =
e^\beta_0= (W/\alpha,Wv/X,0,0)_{\rm{CF}}$ and $n^\beta
=e^\beta_1=(Wv/\alpha,W/X,0,0)_{\rm{CF}}$.  In this we have neglected
rotational effects which will be small for $v_\varphi \ll c$.
Transforming $p^a$ to the coordinate basis of \code{GR1D},
\begin{eqnarray}
p^\beta=p^a e_a^\beta &=&
E^{\rm{FRF}}\left({W \over \alpha} (1+v),{W \over X}(1+v),0,0\right)_{\rm{CF}}.
\end{eqnarray}
An observer at rest in the coordinate frame ($U^\alpha =
(1,0,0,0)_{\rm{CF}}$) then sees the neutrino with energy,
\begin{equation}
\fl E^{\rm{CF}} = -\vec{p} \cdot \vec{U} = - g_{\alpha
  \beta}p^\beta U^\alpha = \alpha^2 E^{\rm{FRF}}{W \over \alpha}(1+v) =
\alpha W (1+v) E^{\rm{FRF}}\, .
\end{equation} 
Noting that (see \cite{MTW}, eq.~25.25), for massless particles
emitted from rest at $r$ and observed by a observer at rest at $r'$,
$\lambda(r) |g_{00}(r)|^{-1/2} = \lambda(r') |g_{00}(r')|^{-1/2}$
implies,
\begin{equation}
{E^{\rm{CF}}(r') \over E^{\rm{CF}}(r)} = {\lambda_{r} \over \lambda_{r'}}=
{|g_{00}(r)|^{1/2} \over |g_{00}(r')|^{1/2}} = {\alpha(r) \over \alpha(r')}\,,
\end{equation}
this is the redshift formula for particles leaving a gravitational well.

\end{document}